\documentclass[final,nomarks]{dmtcs-episciences}
\received{2017-2-5}

\accepted{2017-11-10}

\publicationdetails{19}{2017}{3}{12}{3122}
\usepackage[utf8]{inputenc}
\usepackage{amsmath}
\usepackage{amssymb,latexsym}
\usepackage{cases}
\usepackage{comment}
\usepackage{color}

\title
{Witness structures and immediate snapshot complexes}

\author{Dmitry N. Kozlov}

\affiliation{Department of Mathematics, University of Bremen, Germany}

\newtheorem{theorem}{Theorem}[section]
\newtheorem{df}[theorem]{Definition}
\newtheorem{thm}[theorem]{Theorem} \newtheorem{lemma}[theorem]{Lemma}
\newtheorem{prop}[theorem]{Proposition}
\newtheorem{lm}[theorem]{Lemma}

\newtheorem{rem}[theorem]{Remark}
 \newcommand{\nin}{\noindent}
\newcommand{\pr}{\begin{proof}}

\newcommand{\act}{\text{\rm act}\,}

\newcommand{\cs}{{\mathcal S}}
\newcommand{\codim}{\text{\rm codim}\,}
\newcommand{\csn}{\cs_{\zz_+}}
\newcommand{\da}{\Delta}
\newcommand{\dar}{\downarrow}

\newcommand{\es}{\emptyset}

\newcommand{\last}{\text{\rm last}\,}

\newcommand{\pass}{\text{\rm pass}\,}
\newcommand{\pnt}{round counter }
\newcommand{\pnts}{round counters }
\newcommand{\ra}{\rightarrow}

\newcommand{\st}{\text{\rm{st}}}
\newcommand{\sm}{\setminus}
\newcommand{\smax}{\text{\rm smax}\,}

\newcommand{\supp}{\text{\rm supp}\,}

\newcommand{\tr}{{\bar r}}
\newcommand{\trc}{{\text{\rm Tr}}}
\newcommand{\tq}{{\bar q}}

\newcommand{\wti}{\widetilde}
\newcommand{\zz}{{\mathbb Z}}
\newcommand{\ab}{\allowbreak}

\numberwithin{equation}{section}
\numberwithin{figure}{section}
\numberwithin{table}{section}

\def\pstexInput#1{
}
\begin{document}

\maketitle

\begin{abstract}
In this paper we introduce and study a~new family of combinatorial
simplicial complexes, which we call {\it immediate snapshot
  complexes}. Our construction and terminology are strongly motivated
by theoretical distributed computing, as these complexes are
combinatorial models of the standard protocol complexes associated to
immediate snapshot read/write shared-memory communication model.

In order to define the immediate snapshot complexes we need a~new
combinatorial object, which we call a~{\it witness structure}. These
objects are indexing the simplices in the immediate snapshot
complexes, while a~special operation on them, called {\it ghosting},
describes the combinatorics of taking simplicial boundary. In general,
we develop the theory of witness structures and use it to prove
several combinatorial as well as topological properties of the
immediate snapshot complexes.
\end{abstract}


\section{The motivation for the study of immediate snapshot complexes}
\label{sect:1}

This section contains the motivation for introducing immediate
snapshot complexes, coming from theoretical distributed computing.
The reader interested primarily in the actual family of combinatorial
simplicial complexes, which we define in this paper, and the
mathematics thereof, may skip this section altogether and return to it
at any later point which is deemed to be convenient.

One of the core theoretical models, which is used to understand the
shared-memory communication between a finite number of processes is
the so-called {\it immediate snapshot read/write model}, abbreviated
to IS model. In this model, a~number of processes are set to
communicate by means of a~shared memory.  Each process has an assigned
register, and each process can perform two types of operations: {\it
  write} and {\it snapshot}. The write operation, by definition,
writes the entire state of the process into its assigned register; the
snapshot operation reads the entire memory in one atomic step. The
order in which a process performs these operations is determined by
the distributed protocol, whose execution is asynchronous, satisfying
an additional condition. Namely, we assume that at each turn a group
of processes becomes active. First this group simultaneously writes
its values to the memory, then it simultaneously performs an atomic
read of the entire memory.  The word simultaneously here means that
these processes first all write in some arbitrary order, and this
order does not matter, after which they do snapshots in some order,
and this order does not matter either. In any case, each execution can
be encoded by a sequence of groups of processes which become active at
each turn.

A~closely related computational model is called {\it iterated
  immediate snapshot model}, or IIS model for short. In that model
each process uses fresh memory each time it becomes active again. A
lot of research is dedicated to the IIS model, which has a major
advantage in that the corresponding simplicial model is much
simpler. Furthermore, the solvability in IS is equivalent to the
solvability in IIS, so for these purposes the models are
interchangeable. The questions of round-complexity for solvable tasks
are however very different, as there is a~substantial overhead factor
when simulating IS using IIS. More details on these computational
models, the associated protocol complexes and their equivalence with
each other and with other models can be found in a recent
book~\cite{HKR}, as well as in~\cite{AP,AW,Ha04,HS}.

In this paper we are motivated by the standard full-information
distributed protocols for $n+1$ processes indexed $0,\dots,n$, where
the $k$-th process runs for $r_k$ rounds and then stops. When saying
that the $k$-th process runs for $r_k$ rounds, what we actually mean
is that the $k$-th process becomes active exactly $r_k$ times; each
time it is included into some group of processes, which simultaneously
become active and perform a write followed by a snapshot, as described
above. Let the associate protocol complex be called
$P(r_0,\dots,r_n)$. Our first contribution is to give a rigorous
purely combinatorial definition of $P(r_0,\dots,r_n)$.  To do this, we
introduce new combinatorial objects, which we call {\it witness
  structures} and use them as a~language to define and to analyze this
family of simplicial complexes. The special case $r_0=\dots=r_n=1$
corresponds to the so-called {\it standard chromatic subdivision} of
a~simplex, which was previously considered in the literature, see,
e.g., \cite{subd,HKR,view}. The cases where for some $i$, we have
$r_i\geq 2$ are new. We note, that the iterated standard chromatic
subdivision is related to the above mentioned IIS model, and has been
a~very useful simplicial model to analyze solvability and complexity
of distributed tasks, see, e.g., recent work on the weak symmetry
breaking, \cite{paths,wsb6, wsb12,binom}.


Let us briefly sketch the plan of the article. In Section~\ref{sect:2}
we define witness structures, and a few operations on them, where {\it
  ghosting} is the most important one. That operation encodes the
combinatorics of taking simplicial boundary in the immediate snapshot
complexes. The reason we chose the term {\it ghosting} is because the
operation removes the information of a~set of processes. However,
these processes do not disappear completely, and parts of their
information appears in the total pattern anyway, reflected in the
views of other active processes; so these processes remain like ghosts
hidden in the background. In Section~\ref{sect:3}, we use the language
of witness structures to define the immediate snapshot complexes, and
prove the important Reconstruction Lemma. In Section~\ref{sect:4} we
look at the first properties of the immediate snapshot complexes. In
particular, we prove that they are pure, we look at some enumerative
combinatorics connected to these complexes, and we show how the
standard chromatic subdivision can be seen as an instance of immediate
snapshot complex construction. In Section~\ref{sect:5} we describe a
canonical decomposition of the immediate snapshot complexes, and prove
that topologically, they are pseudomanifolds with boundary. Finally,
in Section~\ref{sect:6}, we explain why our immediate snapshot
complexes provide correct combinatorial model for the protocol
complexes of standard protocols in the immediate snapshot read/write
computation model.

This paper grew out of the first half of author's preprint \cite{K1}.
The second half of \cite{K1} is focused on showing that the immediate
snapshot complexes are homeomorphic to closed balls. Specifically, we
show that there exists a homeomorphism from an~appropriate standard
simplex to the immediate snapshot complex, such that the image of
every subsimplex of the standard simplex is a~subcomplex of the
immediate snapshot complex. This part has already appeared in print,
see~\cite{K2}.


\section{The language of witness structures} \label{sect:2}

\subsection{Some notations} \label{ssect:2.1}

In general, we will not define standard notions related to
combinatorial simplicial complexes, and rather refer to standard
monographs, see~\cite{Hat,book,M84}. However, we do want to fix some
notations.  We let $\zz_+$ denote the set of nonnegative integers
$\{0,1,2,\dots\}$.  For a~nonnegative integer $n$, we shall use $[n]$
to denote the set $\{0,\dots,n\}$, with an~additional convention that
$[-1]=\es$.  For a~finite subset $S\subset\zz_+$, such that $|S|\geq
2$, we let $\smax S$ denote the {\it second} largest element, i.e.,
$\smax S:=\max(S\setminus\{\max S\})$. For a set $S$ and an element
$a$, we set
\[\chi(a,S):=
\begin{cases}
1, & \text{ if } a\in S;\\
0, & \text {otherwise.} 
\end{cases}\]

Whenever $(X_i)_{i=1}^t$ is a~family of topological spaces, we set
$X_I:=\bigcap_{i\in I} X_i$.  Also, when no confusion arises, we
identify one-element sets with that element, and write $p$ instead
of~$\{p\}$.

Furthermore, we need some poset terminology. Recall, that for a poset
$P$ and an element $x\in P$, one sets $P_{<x}:=\{y\in P\,|\,y<x\}$.
In general, a~subset $Q\subseteq P$ is called an~{\it ideal} if for
any $x\in Q$ and $y\in P$, such that $y\leq x$, we have $y\in Q$.
Furthermore, for any subset $A\subseteq P$, we let $I(A,P)$ denote the
set $\{y\in P\,|\,\exists x\in A, \textrm{ such that } y\leq x\}$;
clearly $I(A,P)=\cup_{x\in A}P_{\leq x}$ and is always an ideal.

\subsection{Witness prestructures and structures}

\subsubsection{Definition and examples}

\noindent
We start by defining our main combinatorial gadget.

\begin{df}\label{df:ws}
A~{\bf witness prestructure} is a~finite sequence of pairs of finite
subsets of $\zz_+$, denoted $\sigma=((W_0,G_0),\dots,(W_t,G_t))$, with
$t\geq 0$, satisfying the following conditions:
\begin{enumerate}
\item[(P1)] $W_i,G_i\subseteq W_0$, for all $i=1,\dots,t$;
\item[(P2)] $G_i\cap G_j=\emptyset$, for all $0\leq i<j\leq t$;
\item[(P3)] $G_i\cap W_j=\emptyset$, for all $0\leq i\leq j\leq t$;
\item[(P4)] $(W_i,G_i)\neq (\es,\es)$, for all $1\leq i\leq t$.
\end{enumerate}

\noindent
A witness prestructure is called {\bf stable} if in addition the following
condition is satisfied:
\begin{enumerate}
\item[(S)] if $t\geq 1$, then $W_t\neq\emptyset$.
\end{enumerate}

\noindent
A {\bf witness structure} is a witness prestructure satisfying the
following strengthening of conditions (S) and (P4):
\begin{enumerate}
\item[(W)] the subsets $W_1,\dots,W_t$ are all nonempty.
\end{enumerate}
\end{df}

\begin{figure}[hbt]
\[
\begin{array}{|c|c|c|c|c|}
\hline
W_0 & W_1 & W_2 & \dots & W_t \\ \hline
G_0 & G_1 & G_2 & \dots & G_t \\ 
\hline
\end{array}\]
\caption{Table presentation of a witness (pre)structure.}
\label{table:ws}
\end{figure}

It is often useful to depict a~witness prestructure in form of
a~table, see Figure~\ref{table:ws}. The intuition behind this notation
is as follows. The first column is essentially for book-keeping
purposes only, as the set $W_0\cup G_0$ tells us which processes are
present at all. The set $G_0$ contains the processes which are not
active and which have not been witnessed by anybody. For $k\geq 1$ the
$k$-th column tells us about the information we have at $k$-th
turn. The set $W_k$ describes the processes which are active at this
turn, while the set $G_k$ describes the processes which became the
ghosts at this turn. The reader should note that the word {\it turn}
does not mean the same here as in the execution of the protocol. It
can happen that such a large number of processes becomes ghosts that
there are several turns in the original protocol where the active
processes do not witness anything. Rather, the columns of the table on
Figure~\ref{table:ws} should be thought as the {\it witnessing turns}.

The axioms in Definition~\ref{df:ws} can now be given intuitive
interpretation. The axiom (P1) says that $W_0$ records processes which
are active at some point. Note that we allow processes in 
\[W_0\sm(W_1\cup\dots\cup W_t\cup G_0\cup\dots\cup G_t),\] 
these correspond to processes which are only required to execute $0$
times but still declare themselves active. Axiom (P2) says that each
process can become a ghost at at most one point in time. Axiom (P3)
says that once a process has become a ghost it cannot be listed as
active at a~later point in time. Axiom (W) (which subsumes (P4) and
(S)) says that at each witnessing step we have some process which is
active.

The 3 possibilities provided by Definition~\ref{df:ws} are illustrated
on Figure~\ref{table:wsex}. On that figure, the witness structure
$\sigma_3$ encodes the following view of process number $1$: the
process has been active once, at which point it has witnessed
processes $2$ and $3$, and it has also witnessed that process $2$ has
witnessed process $0$ at some earlier stage of the execution. The
process $1$ did not witness process $4$ and it also did not witness
anyone who has witnessed $4$; however the witness structure still
records the fact that process $4$ exists. The witness structure
$\sigma_3$ will index a~vertex in the corresponding immediate snapshot
simplicial complex.

\begin{figure}[hbt]
\[
\begin{array}{|c|c|c|c|c|}
\hline
[4] & 1& \emptyset & 2,3 & \emptyset \\ \hline
5       &       \emptyset & 1 & 4 & 3  \\ 
\hline
\end{array}
\quad\quad
\begin{array}{|c|c|c|c|c|}
\hline
[3] & \emptyset & 2 & \emptyset & 1 \\ \hline
4       &  0  &     \emptyset & 2 & 3  \\ 
\hline
\end{array}
\quad\quad
\begin{array}{|c|c|c|}
\hline
[3] & 2 & 1 \\ \hline
4 & 0 & 2,3 \\ 
\hline
\end{array}\]
\caption{A witness prestructure $\sigma_1$, a stable witness
  prestructure $\sigma_2$, and a~witness structure $\sigma_3$.}
\label{table:wsex}
\end{figure}

Note, that every witness prestructure with $t=0$ is a~witness
structure. On the other hand, if $W_0=\emptyset$, then conditions (P1)
and (P4) imply that $t=0$. In this case, only the set $G_0$ carries any
information, and we call this witness structure {\it empty}.

Let us remark here that the main objects we would like to study are
the witness structures. The reason being that they index simplices in
the immediate snapshot complexes. In order to reflect the incidence
structure of the immediate snapshot complexes we need to study the
so-called {\it ghosting operation} on the witness structures, including
the effect of the iteration of ghosting. It is technically much easier
to do that in the more general context of prestructures, which is the
main reason why we introduce these objects.

\subsubsection{Data associated to witness prestructures}

\begin{df}
Let $\sigma=((W_0,G_0),\dots,(W_t,G_t))$ be an arbitrary witness
prestructure. We define the following data associated to $\sigma$:
\begin{itemize}
\item the set $W_0\cup G_0$ is called the {\bf support} of $\sigma$ and is
denoted by $\supp\sigma$;
\item the {\bf ghost set} of $\sigma$ is the set
  $G(\sigma):=G_0\cup\dots\cup G_t$;
\item the {\bf active set} of $\sigma$ is the complement of the ghost set
\[A(\sigma):=\supp(\sigma)\setminus G(\sigma)=W_0\setminus(G_1\cup\dots\cup G_t);\]
\item the {\bf dimension} of $\sigma$ is
  \[\dim\sigma:=|A(\sigma)|-1=|W_0|-|G_1|-\dots-|G_t|-1.\]
\end{itemize}
\end{df}

For the examples on Figure~\ref{table:wsex} we get
$\supp\sigma_1=[5]$, $\supp\sigma_2=\supp\sigma_3=[4]$,
$G(\sigma_1)=\{1,3,4,5\}$, $G(\sigma_2)=G(\sigma_3)=\{0,2,3,4\}$,
$A(\sigma_1)=\{0,2\}$, $A(\sigma_2)=A(\sigma_3)=\{1\}$,
$\dim\sigma_1=1$, and $\dim\sigma_2=\dim\sigma_3=0$.

By definition, the dimension of a~witness prestructure $\sigma$ is between $-1$ and $|\supp\sigma|-1$. Let us analyze the witness structures which have special dimension. To start with, if $\dim(\sigma)=-1$, then $|A(\sigma)|=0$, i.e., $A(\sigma)=\emptyset$, hence $W_0=G_1\cup\dots\cup G_t$. On the other hand, if $t\geq 1$, by (P3) we have
\[W_t\cap(G_1\cup\dots\cup G_t)=\emptyset,\] and by (P1) we have $W_0\supseteq W_t$,
implying that $W_t=\emptyset$. Hence, by (P1) and (P4), the only witness structures of
dimension $-1$ are empty, i.e., of the form $\sigma=((\emptyset,G_0))$.

Note, that by (P3), the set $W_k$ is in general disjoint from
$G_0\cup\dots\cup G_k$, for all $0\leq k\leq t$, so we have
\[W_k\subseteq A(\sigma)\cup G_{k+1}\cup\dots\cup G_t.\]
Furthermore, it is easy to characterize all witness structures
$\sigma$ of dimension~$0$.  In this case, we have $|A(\sigma)|=1$. We
let $\sigma=((W_0,G_0),\dots,(W_t,G_t))$ and let $p$ denote the unique
element of $A(\sigma)$, then $\sigma$ has dimension~$0$ if and only if
\[W_k\subseteq\{p\}\cup G_{k+1}\cup\dots\cup G_t,
\text{ for all } k=0,\dots,t.\] 
In particular, we must of course have $W_t=\{p\}$, and we shall call
$p$ the {\it color} of the witness structure~$\sigma$.

At the opposite extreme, a~witness structure
$\sigma=((W_0,G_0),\dots,(W_t,G_t))$ has dimension $|\supp\sigma|-1$
if and only if $G_0=\dots=G_t=\emptyset$. In such a~situation, we
shall frequently use the short-hand notation
$\sigma=(W_0,W_1,\dots,W_t)$.

\subsubsection{Traces and alternative definition of witness structures} 

\noindent
For brevity of some formulas, we set $W_{-1}:=W_0\cup G_0=
\supp\sigma$.

\begin{df}\label{df:trc}
For a~prestructure $\sigma$ and an arbitrary $p\in\supp\sigma$, we set
\[\trc(p,\sigma):=\{0\leq i\leq t\,|\,p\in W_i\cup G_i\},\] 
and call it the {\bf trace} of~$p$. Furthermore, for all
$p\in\supp\sigma$, we set 
\[\last(p,\sigma):=\max\{-1\leq i\leq t\,|\,p\in W_i\}.\]
\end{df}

\nin When the choice of $\sigma$ is unambiguous, we shall simply write
$\trc(p)$ and $\last(p)$. Note that $0\in\trc(p)$, hence
$\trc(p)\neq\emptyset$. Note furthermore, that if $p\in A(\sigma)$,
then $\trc(p)=\{0\leq i\leq t\,|\,p\in W_i\}$, while $p\in G(\sigma)$
implies $\trc(p)\sm\max\trc(p)=\{0\leq i\leq t\,| \allowbreak\,p\in
W_i\}$ and $p\in G_{\max\trc(p)}$.

To get a~better grasp on the witness structures, as well as operations
in them, the following alternative definition, which uses traces, is
often of value.

\begin{df}\label{df:trws}
Let $A$ and $G$ be disjoint finite subsets of $\zz_+$, and let
$\{\trc(p)\}_{p\in A\cup G}$ be a~family of finite subsets of $\zz_+$.
Set $t:=\max\bigcup_{p\in A\cup G}\trc(p)$. The triple
$(A,G,\{\trc(p)\}_{p\in A\cup G})$ is called a~{\bf witness
  prestructure} if the following conditions are satisfied:
\begin{itemize}
\item[(T1)] $0\in\trc(p)$, for all $p\in A\cup G$;
\item[(T2)] $\bigcup_{p\in A\cup G}\trc(p)=[t]$.
\end{itemize}

\noindent
A witness prestructure is called {\bf stable} if it satisfies
an~additional condition:
\begin{itemize}
\item[(TS)] if $A=\emptyset$, then $\trc(p)=\{0\}$, for all $p\in G$,
else 
$\max\bigcup_{p\in A}\trc(p)=t$.
\end{itemize}
 A~stable witness prestructure is
called {\bf witness structure} if the following strengthening of
Condition (TS) is satisfied:
\begin{enumerate}
\item[(TW)] {\it for all $1\leq k\leq t$ either there exists $p\in A$
  such that $k\in\trc(p)$, or there exists $p\in G$ such that
  $k\in\trc(p)\sm\max\trc(p)$.}
\end{enumerate}
\end{df}

We shall call the form of the presentation of the witness prestructure
as a~triple $(A,G,\{\trc(p)\}_{p\in A\cup G})$ its {\it trace form}.

Intuitively, each trace set $\trc(p)$ records the turns in which process $p$ is involved, either actively or as a ghost. The axiom (T1) means that each process is recorded at step $0$ which only serves the bookkeeping purpose of telling us which processes do we have. Axiom (T2) means that some process is recorded at each step, so there are no ``skips''. Axioms (TS) and (TW) are very close to our previous axioms (S) and (W). Axiom (TS) says that some non-ghost process is active at the last turn, whereas axiom (TW) says that this is the case at every turn.

\begin{prop}
There is a~natural bijection between the objects described by
the Definitions~\ref{df:ws} and~\ref{df:trws}.
\end{prop}
\pr The translation between the two descriptions is as follows.
First, assume $\sigma=((W_0,G_0),\allowbreak \dots, \allowbreak
(W_t,G_t))$ is a~witness prestructure according to
Definition~\ref{df:ws}. Set $A:=A(\sigma)$, $G:=G(\sigma)$, and for
each $p\in A\cup G$, set $\trc(p)$ to be the trace of $p$ as given by
Defi\-nition~\ref{df:trc}. The condition (T1) is then satisfied since
$\supp\sigma=A(\sigma)\cup G(\sigma)=W_0\cup G_0$. The condition~(T2)
is implied by (P4). If $\sigma$ is a~stable prestructure, then
$\max_{p\in A}\last(p)=t$, hence the condition (TS) is
satisfied. Finally, if $\sigma$ is a~witness structure, then
$W_i\neq\emptyset$ for all $i=1,\dots,t$. Assume $p\in W_i$. If $p\in
A(\sigma)$, then $i\in\trc(p)$, else $p\in G(\sigma)$, and
$i\in\trc(p)\sm\max\trc(p)$. In any case, the condition (TW) is
satisfied.

Reversely, assume we are given a~triple $(A,G,\{\trc(p)\}_{p\in A\cup
  G})$ as in Definition~\ref{df:trws}. We set 
\[t:=\max\bigcup_{p\in A\cup G}\trc(p),\] 
and for all $0\leq k\leq t$, we set
\[G_k:=\{p\in G\,|\,k=\max\trc(p)\},\]
\[W_k:=\{p\in A\cup G\,|\,k\in\trc(p)\}\setminus G_k.\]

\nin In particular, $G_0=\{p\in G\,|\,\trc(p)=\{0\}\}$, and by (T1) we
have $W_0=(A\cup G)\sm G_0$. It follows that $W_i,G_i\subseteq W_0$,
for all $i=1,\dots,t$, and (P1) is satisfied. Furthermore, we have
$W_i\cap G_i=\emptyset$, and $G_i\cap G_j=\emptyset$, for $i\neq j$,
by construction, so (P2) is satisfied. Still by construction, we have
$G_i\cap W_j=\{p\in G\,|\,i=\max\trc(p),\,j\in\trc(p)\}$, which is
clearly empty when $i<j$.

The condition (TS) implies (S), since it implies that there exists an element
$p\in A$, such that $\max\trc(p)=t$, so $p\in W_t$. Finally, (TW)
implies (W), since both $p\in A$, $k\in\trc(p)$ and $p\in G$,
$k\in\trc(p)\sm\max\trc(p)$, imply $p\in W_k$. We leave it to the
reader to verify that the translations described above are inverses of
each other.  \end{proof}

\vspace{5pt}

\noindent
For example, for the prestructure $\sigma_1$ on
Figure~\ref{table:wsex}, we get $\trc(0)=\trc(5)=\{0\}$,
$\trc(1)=[2]$, $\trc(2)=\trc(4)=\{0,3\}$, and $\trc(3)=\{0,3,4\}$.

\subsection{Operations on witness prestructures}

\subsubsection{Stabilization of witness prestructures}

\noindent
In order to gain a~more clear understanding of the composition of the
ghosting operation with itself, we will break that operation into two
parts: stabilization modulo a~set and taking canonical form. We start
by noting, that any witness prestructure can be made stable using
the following operation.

\begin{df} \label{df:st}
Let $\sigma=(A,G,\{\trc(p)\}_{p\in A\cup G})$ be a~witness
prestructure. We set
\[q:=\begin{cases}
\max\bigcap_{p\in A}\trc(p),&\textrm{ if }A\neq\es,\\
0,&\textrm{ otherwise.}
\end{cases}\]
The {\bf stabilization} of $\sigma$ is the witness
prestructure $\st(\sigma)$ whose trace form is $(A,G,\{\trc(p)\cap
[q]\}_{p\in A\cup G})$.
\end{df}

The reader will notice that we use the trace form definition of the
witness prestructure to define stabilization. The equivalent
formulation for the table form of the witness prestructure is of
course possible, but unfortunately is more involved. The stabilization
is done by first finding the rightmost column with an active process,
truncating at this column, and then for each ghost process which
appeared in the truncated part we find the rightmost column in which
it appears and move it in this column from the active state to the
ghost state. An~example is shown on Figure~\ref{table:ws4}, where
$A=\{0,4\}$, $G=\{1,2,3\}$, and $q=2$.

\begin{figure}[hbt]
\[
\begin{array}{|c|c|c|c|c|c|c|}
\hline
[4] & 1   & 0,3,4 & 2,3 & 1 & 1 & \es  \\ \hline
\es       & \es & \es   & \es & 3 & 2 & 1 \\ 
\hline
\end{array}
\longrightarrow
\begin{array}{|c|c|c|}
\hline
[4] & 1 & 0,3,4 \\ \hline
\es & \es & \es \\ 
\hline
\end{array}
\longrightarrow
\begin{array}{|c|c|c|}
\hline
0,1,3,4 & \es & 0,4 \\ \hline
      2 & 1   &  3\\ 
\hline
\end{array}\]
\caption{Stabilizing a witness prestructure.}
\label{table:ws4}
\end{figure}

\begin{prop} \label{prop:stall0}
For an arbitrary witness prestructure $\sigma$, the witness
prestructure $\st(\sigma)$ is well-defined and stable. These two
prestructures have the same support, dimension, ghost and active
sets. Furthermore, we have $\sigma=\st(\sigma)$ if and only if
$\sigma$ is stable.
\end{prop}
\pr Both (T1) and (T2) are immediate, since we are simply restricting
traces to the set $[q]$, where $q\geq 0$. The stability condition (TS)
follows from the choice of $q$. \end{proof}

\subsubsection{Canonical form of a stable witness prestructure} 

\noindent
As the next step, any stable witness prestructure can be turned into
a~witness structure by means of the following operation.

\begin{df}\label{df:cform}
Assume $\sigma=((W_0,G_0),\dots,(W_t,G_t))$ is an arbitrary stable
witness prestructure. Set $q:=|\{1\leq i\leq t\,|\,
W_i\neq\emptyset\}|$.  Pick $0=i_0<i_1<\dots<i_q=t$, such that
\[\{i_1,\dots,i_q\}=\{1\leq i\leq t\,|\,W_i\neq\emptyset\}.\]  We define
the witness structure $C(\sigma)=((W_0,G_0),\ab(\wti W_1, \wti
G_1),\allowbreak \dots,\ab(\wti W_q,\wti G_q))$, which is called the {\bf
  canonical form} of~$\sigma$, by setting
\begin{equation}\label{eq:canon}
\wti W_k:=W_{i_k},\quad 
\wti G_k:=G_{i_{k-1}+1}\cup\dots\cup G_{i_k},\text{ for all } k=1,\dots,q,
\end{equation}
\end{df}

A graphical way of thinking about obtaining the canonical form, is that all the columns with the empty set in the upper box get ``merged to the right''. Since we assume that the upper box of the last column is not empty, this is well-defined for the entire table. 

The reason for the asymmetry when merging to the right is because the time of the turns increases from left to right, so when some witnesses disappear, their information will be recovered at later turns. The construction in Definition~\ref{df:cform} is illustrated by Figure~\ref{table:ws2}. 

\begin{figure}[hbt]
\[
\begin{array}{|c|c|c|c|c|}
\hline
[3] & \emptyset & 2         & \emptyset & 1 \\ \hline
4   &  0        & \emptyset & 2         & 3  \\ 
\hline
\end{array}
\longrightarrow
\begin{array}{|c|c|c|}
\hline
[3] & 2 & 1 \\ \hline
4   & 0 & 2,3 \\ 
\hline
\end{array}\]
\caption{A stable witness prestructure and its canonical form.}
\label{table:ws2}
\end{figure}

\begin{prop} \label{prop:call}
Assume $\sigma$ is an arbitrary stable witness prestructure.
\begin{itemize}
\item[(a)] The canonical form of $\sigma$ is a~well-defined witness
  structure.
\item[(b)] We have $C(\sigma)=\sigma$ if and only if $\sigma$ is
  itself a~witness structure.
\item[(c)] We have  $\supp(C(\sigma))=\supp(\sigma)$,
$A(\sigma)=A(C(\sigma))$, $G(\sigma)=G(C(\sigma))$, and finally
$\dim(\sigma)=\ab\dim(C(\sigma))$.
\end{itemize}
\end{prop}

\pr Assume $\sigma=((W_0,G_0),\dots,(W_t,G_t))$, $q$ and
$i_1,\dots,i_q$ as in the Definition~\ref{df:cform}, and
$C(\sigma)=((W_0,G_0),(\wti W_1,\wti G_1),\dots,(\wti W_q,\wti G_q))$. 

To prove (a) note first that all the sets involved are finite subsets 
of $\zz_+$. Conditions (P1) and (P2) for $C(\sigma)$ follow immediately 
from the corresponding conditions on~$\sigma$. To see (P3), pick some 
$p\in\wti G_k$. Then there exists a~unique $j$, such that 
$i_{k-1}<j\leq i_k$ and $p\in G_j$. Then $p\notin W_j\cup\dots\cup W_t$, 
but $W_j\cup\dots\cup W_t=W_{i_k}\cup\dots\cup W_{i_q}$, hence 
$p\notin\wti W_k\cup\dots\cup\wti W_q$. Finally, to see (W) note that
$W_{i_k}\neq\emptyset$ for all $k=1,\dots,q$, hence $\wti W_k\neq\emptyset$. 

To prove (b) note that if $\sigma$ is a~witness structure, then 
$W_1,\dots,W_t\neq\emptyset$, hence $q=t$, $i_k=k$, for $k=1,\dots,t$. 
It follows that $\wti W_k=W_k$, $\wti G_k=G_k$, for all $k=1,\dots,t$. 
Reversely, assume $C(\sigma)=\sigma$, then $q=t$, hence $i_k=k$, 
for all $k=1,\dots,t$, implying $W_1,\dots,W_t\neq\emptyset$.

To prove (c) note that the first pair of sets in $\sigma$ and in $C(\sigma)$ is 
the same, hence $\supp(C(\sigma))=\supp(\sigma)$. By \eqref{eq:canon} we have 
$\wti W_1\cup\dots\cup\wti W_q=W_1\cup\dots\cup W_t$, and $\wti
G_1\cup\dots\cup\wti G_q=\ab G_1\cup\dots\cup G_t$, hence
$A(\sigma)=A(C(\sigma))$. The other two equalities follow. 
\end{proof}
\vspace{5pt}


\subsubsection{Stabilization of witness prestructures modulo a set of processes} 

\noindent
The Definition~\ref{df:st}  can be generalized as follows.

\begin{df} \label{df:st2}
Let $\sigma=(A,G,\{\trc(p)\}_{p\in A\cup G})$ be a~witness
prestructure, and $S\subseteq A$. If $S\subset A$, set $q:=\max
\bigcup_{p\in A\sm S}\trc(p)$, else set $q:=0$. The {\bf
  stabilization} of $\sigma$ {\bf modulo} $S$ is the witness
prestructure $\st_S(\sigma)$ whose trace form is $(A\setminus S,G\cup
S,\{\trc(p)\cap[q]\}_{p\in A\cup G})$.
\end{df}

Intuitively speaking, stabilization modulo $S$ is just like the usual stabilization defined above, with the addition that the processes from $S$ are not considered to be active, so the truncation happens at the rightmost column which contains an active process which does not belong to~$S$.

The following three properties can be taken as a~recursive alternative
to Definition~\ref{df:st2}.
\begin{enumerate}
\item[(1)] If $t=0$, then $\st_S(\sigma)=((W_0\setminus S,G_0\cup S))$.
\item[(2)] If $t\geq 1$ and $W_t\subseteq S$, then 
\[\st_S(\sigma)=\st_{S\cup G_t}(((W_0,G_0),\dots,(W_{t-1},G_{t-1}))).\]
\item[(3)] If $t\geq 1$ and $W_t\not\subseteq S$, then the trace form
of $\st_S(\sigma)$ is $(A(\sigma)\setminus S,\ab G(\sigma)\cup S,\ab 
\{\trc(p)\}_{p\in A\cup G})$.
\end{enumerate}

Assume now that $\st_S(\sigma)=((\wti W_0,\wti G_0),\dots,(\wti
W_q,\wti G_q))$.  By Definition~\ref{df:st2} we have $\wti W_i\cup\wti
G_i=W_i\cup G_i$, and $W_i\supseteq\wti W_i$, for all $0\leq i\leq q$.
Hence, for some sets $J_0,\dots,J_q$ we have
\begin{equation}\label{eq:tfst}
\st_S(\sigma)=((W_0\sm J_0,G_0\cup J_0),\dots,(W_q\sm J_q,G_q\cup J_q)).
\end{equation}
The sets $J_i$ can be explicitly described by the following formula:
\[J_i:=(S\cup G(\sigma))\cap (W_i\setminus(W_{i+1}\cup\dots\cup W_q\cup 
G_{i+1}\cup\dots\cup G_q)).\]

Note that unlike the usual stabilization, the witness prestructure $\st_S(\sigma)$ might be different from $\sigma$ even when $\sigma$ is a~witness structure.

\begin{prop}\label{prop:stall}
Assume as before that we are given a~witness prestructure $\sigma$,
and $S\subset A(\sigma)$. Then, the witness prestructure
$\st_S(\sigma)$ is well-defined and stable. It satisfies the following
properties:
\begin{enumerate}
\item[(1)] $\supp(\st_S(\sigma))=\supp\sigma$;
\item[(2)] $G(\st_S(\sigma))=G(\sigma)\cup S$;
\item[(3)] $A(\st_S(\sigma))=A(\sigma)\sm S$;
\item[(4)] $\dim\st_S(\sigma)=\dim\sigma-|S|$.
\end{enumerate}
\end{prop}
\pr Clearly, the conditions (T1) and (T2) are still satisfied, so the
witness prestructure $\st_S(\sigma)$ is well-defined. Using
an~argument verbatim to the proof of the
Proposition~\ref{prop:stall0}, we conclude that it is stable due to
the choice of~$q$. The identities $(2)$ and $(3)$ are integral parts
of Definition~\ref{df:st2}, and $(1)$ and $(4)$ are direct
consequences.\end{proof}

\vspace{5pt}

\noindent
The following property of the stabilization will be very useful later on.

\begin{prop} 
\label{prop:stst}
Assume $\sigma$ is a witness prestructure, and $S,T\subseteq A(\sigma)$, 
such that $S\cap T=\es$. Then we have
\begin{equation}\label{eq:stst}
\st_T(\st_S(\sigma))=\st_{S\cup T}(\sigma).
\end{equation}
\end{prop}

\pr Assume $\sigma=(A,G,\{\trc(p)\}_{p\in A\cup G})$, and set
$\sigma':=\st_T(\st_S(\sigma))$, $\sigma'':=\st_{S\cup T}(\sigma)$.
To show that $\sigma'=\sigma''$ we compare their trace forms.  To
start with, by Definition~\ref{df:st2} we have
$\supp\sigma'=\supp\sigma$ and $\supp\sigma''=\supp\sigma$. Furthermore,
$A(\sigma'')=A\setminus(S\cup T)$, and $A(\sigma')=A(\st_S(\sigma))
\setminus T=(A\setminus S)\setminus T$, hence $A(\sigma')=A(\sigma'')$
and $G(\sigma')=G(\sigma'')$.

Finally, both in $\sigma'$ as well as in $\sigma''$ the traces of
elements from $A\cup G$ are truncated at the index $\max\bigcup_{p\in
  A\sm(S\cup T)}\trc(p)$.  \end{proof}


\subsubsection{Ghosting operation on the witness structures} 

\vspace{5pt}

\noindent
We are now ready to define the main operation on witness structures.

\begin{df}\label{df:go}
For an arbitrary witness structure $\sigma$, and an~arbitrary
$S\subseteq A(\sigma)$, we define
$\Gamma_S(\sigma):=C(\st_S(\sigma))$. We say that $\Gamma_S(\sigma)$
is obtained from $\sigma$ {\bf by ghosting~$S$}.
\end{df}
\noindent The ghosting operation is illustrated on Figure~\ref{table:ws3}. In the simplicial context, this figure shows how to find one of the end vertices of the edge indexed by the witness structure on the left.

\begin{figure}[hbt]
\[
\begin{array}{|c|c|c|c|c|}
\hline
[3]   & 3 & 1,2 & 3 & 3  \\ \hline
\emptyset & \emptyset &\es & 0 & 1 \\ 
\hline
\end{array}
\longrightarrow
\begin{array}{|c|c|c|}
\hline
1,2,3 & \es& 2 \\ \hline
0 & 3& 1 \\ 
\hline
\end{array}
\longrightarrow
\begin{array}{|c|c|}
\hline
1,2,3 & 2 \\ \hline
0 & 1,3 \\ 
\hline
\end{array}\]
\caption{Ghosting a witness structure with respect to $S=\{3\}$; here
  the first arrow represents stabilizing this structure with respect
  to~$S$.}
\label{table:ws3}
\end{figure}

Clearly, we have $\Gamma_\es(\sigma)=\sigma$. If $S=\{p\}$, we shall
simply write $\Gamma_p(\sigma)$. In this case we are ghosting a~single
element, and though the situation is not quite straightforward,
several special cases can be formulated in a~simpler manner.

Let $l:=\last(p)$. If $|W_l|\geq 2$, then the situation is much
simpler indeed. In this case $J_i=\emptyset$, for all $i\neq l$, 
while $J_l=\{p\}$. Accordingly, we get 
\begin{multline*}
\Gamma_p(\sigma)=((W_0,G_0),\dots,(W_{l-1},G_{l-1}),
(W_l\setminus\{p\},G_l\cup\{p\}),(W_{l+1},G_{l+1}),\dots,(W_t,G_t)).
\end{multline*}

The situation is slightly more complex if $|W_l|=1$, i.e.,
$W_l=\{p\}$.  Assume that $l\leq t-1$. Then, we still have $J_i=\es$,
for all $i\neq l$, and $J_l=\{p\}$.  The difference now is that
\[\st_S(\sigma)=((W_0,G_0),\dots,(W_{l-1},G_{l-1}),
(\emptyset,G_l\cup\{p\}),(W_{l+1},G_{l+1}),\dots,(W_t,G_t))\] is now
only a~stable witness prestructure, so in this case we get
\[
\Gamma_p(\sigma)=((W_0,G_0),\dots,(W_{l-1},G_{l-1}),
(W_{l+1},G_l\cup\{p\}\cup G_{l+1}),\dots,(W_t,G_t)).
\]
Once $l=t$, i.e., $W_t=\{p\}$, we will need the full generality of the
Definition~\ref{df:go}.

The situation is similar if $|S|\geq 2$. For each element $s\in S$ we
set $l(s):=\last(s)$. As long as each $W_{l(S)}$ contains elements
outside of $S$, all that happens is that each element $s\in S$ gets
moved from $W_{l(S)}$ to $G_{l(S)}$. Once this is not true, a more
complex construction is needed.
 
\begin{prop}\label{prop:gall}
Assume we are given a witness structure $\sigma=(A,G,\{\trc(p)\}_{p\in
  A\cup G})$, and an~arbitrary $S\subseteq A$. The
construction in Definition~\ref{df:go} is well-defined, and yields
a~witness structure $\Gamma_S(\sigma)$, satisfying the following
properties:
\begin{enumerate}
\item[(1)] $\supp(\Gamma_S(\sigma))=A\cup G$;
\item[(2)] $G(\Gamma_S(\sigma))=G\cup S$;
\item[(3)] $A(\Gamma_S(\sigma))=A\sm S$;
\item[(4)] $\dim\Gamma_S(\sigma)=\dim\sigma-|S|$.
\end{enumerate}
\end{prop}
\pr All equalities follow from the Propositions~\ref{prop:call}
and~\ref{prop:stall}.  
\end{proof}

\begin{rem}\label{rem:tr}
For future reference we make the following observation.  Let
$\sigma=((W_0,G_0),\ab\dots,\ab(W_t,G_t))$ be a~witness structure, and
assume $p,q\in\supp\sigma$, $p\neq q$. Clearly, we have
$|\trc(q,\Gamma_p(\sigma))|\leq |\trc(q,\sigma)|$. Furthermore, if
$W_t\neq\{p\}$ we get an equality
$|\trc(q,\Gamma_p(\sigma))|=|\trc(q,\sigma)|$, for all~$q$.
\end{rem}


\begin{lemma} \label{lm:1}
Assume $\sigma$ is a~stable prestructure, and $S\subseteq A(\sigma)$,
then we have $C(\st_S(\sigma))=C(\st_S(C(\sigma)))$, or expressed
functorially $C\circ\st_S\circ C=C\circ\st_S$.
\end{lemma}
\pr Assume $\sigma=((W_0,G_0),\dots,(W_t,G_t))$. We first describe the
witness structure $C(\st_S(C(\sigma)))$. By Definition~\ref{df:st}, we
have $C(\sigma)=((W_{i_0},\wti G_{i_0}),\ab \dots,(W_{i_q},\wti
G_{i_q}))$, where $\wti G_{i_k}=
\ab\bigcup\limits_{\alpha=i_{k-1}+1}^{i_k} G_\alpha$, for all
$k=0,\dots,q$, and the indices $q$ and $0=i_0<i_1<\dots<i_q=t$ are
chosen appropriately.

Set $\wti S:=S\cup G(\sigma)$, set $r:=\max\{0\leq k\leq q\,|\,
W_{i_k}\not\subseteq\wti S\}$, and set $J_k:=\wti S\cap(W_{i_k}\setminus
\bigcup\limits_{\alpha=k+1}^q(W_{i_\alpha}\cup\widetilde
G_{i_\alpha}))$, for $0\leq k \leq r$. Then
\[\st_S(C(\sigma))=((W_{i_0}\setminus J_0,\wti G_{i_0}\cup J_0),
\dots,(W_{i_r}\setminus J_r,\wti G_{i_r}\cup J_r)).\]

On the other hand, we have $i_r=\max\{0\leq j\leq
t\,|\,W_j\not\subseteq\wti S\}$ and $J_k=\wti S\cap(W_{i_k}
\setminus\bigcup\limits_{\alpha=i_k+1}^t (W_\alpha\cup G_\alpha))$,
for $0\leq k\leq r$, since $W_j=\es$ whenever
$j\not\in\{i_0,\dots,i_q\}$, and $\widetilde
G_{i_k}=\bigcup\limits_{\alpha=i_{k-1}+1}^{i_k} G_\alpha$, for all
$k=0,\dots,q$. It follows, that we have
$\st_S(\sigma)=((W'_0,G'_0),(W'_1,G'_1),\dots,(W'_{i_r},G'_{i_r}))$,
where
\begin{equation}\label{eq:wgj}
(W'_j,G'_j)=\begin{cases}
(W_{i_k}\setminus J_k,G_{i_k}\cup J_k,), & \text{ if } j=i_k, \text{ for some }
0\leq k\leq r, \\
(\es,G_j), & \text{ otherwise}.
\end{cases} 
\end{equation}

Set $d:=|\{0\leq k\leq r\,|\,W_{i_k}\setminus J_k\neq\es\}|$, 
then~\eqref{eq:wgj} implies that we also have 
\[d=|\{0\leq k\leq i_r\,|\,W'_k\neq\es\}|.\] This means that 
$C(\st_S(\sigma))$ and $C(\st_S(C(\sigma)))$ have the same length.

For the appropriate choice of $0=a(0)<a(1)<\dots<a(d)=r$ we have 
\[\{a(0),\dots,a(d)\}=\{0\leq k\leq r\,|\,W_{i_k}\setminus J_k\neq\es\}.\]
Assume $C(\st_S(C(\sigma)))=((V_0,H_0),\dots,(V_d,H_d))$, then we have
$V_k=W_{i_{a(k)}}\setminus J_{a(k)}$,
\begin{equation}\label{eq:hk}
H_k=\bigcup\limits_{\alpha=a(k-1)+1}^{a(k)}(\wti G_{i_\alpha}\cup J_\alpha)=
\bigcup\limits_{\alpha=a(k-1)+1}^{a(k)}\wti G_{i_\alpha}\cup
\bigcup\limits_{\alpha=a(k-1)+1}^{a(k)} J_\alpha,
\end{equation}
for $0\leq k\leq d$. 

Assume now that $C(\st_S(\sigma))=((V'_0,H'_0),\dots,(V'_d,H'_d))$. Note that
\[\{i_{a(0)},\dots,i_{a(d)}\}=\{0\leq k\leq i_r\,|\,W'_k\neq\es\},\] 
hence, for $0\leq k\leq d$, we get
$V'_k=W'_{i_{a(k)}}=W_{i_{a(k)}}\setminus J_{a(k)}$, and 
\[H'_k=\bigcup\limits_{\alpha=i_{a(k-1)}+1}^{i_{a(k)}}G'_\alpha=
\bigcup\limits_{\alpha=i_{a(k-1)}+1}^{i_{a(k)}}G_\alpha\cup
\bigcup\limits_{\alpha=a(k-1)+1}^{a(k)}J_\alpha,\]
where the last equality is a consequence of~\eqref{eq:wgj}. 
Combining the identity
\[\bigcup\limits_{\alpha=a(k-1)+1}^{a(k)}\wti G_{i_\alpha}=
\bigcup\limits_{\alpha=a(k-1)+1}^{a(k)}
\bigcup_{\beta=i_{\alpha-1}+1}^{i_\alpha} G_\beta=
\bigcup\limits_{\beta=i_{a(k-1)}+1}^{i_{a(k)}}G_\beta\]
with~\eqref{eq:hk}, we see that $H_k=H'_k$, for all $0\leq k\leq d$.
\end{proof}

The next theorem is the pinnacle of our efforts. It allows us to ghost processes in an arbitrary order. A~direct proof of this fact would be possible, but technically involved and unilluminatig. This is reason, why we choose to define stable prestructures, stabilization and canonical form. Having worked out the necessary properties of these, we can now give an extremely short and purely symbolic argument.

\begin{thm} 
\label{thm:gg}
Assume $\sigma$ is a witness structure, and $S,T\subseteq A(\sigma)$, 
such that $S\cap T=\emptyset$. Then we have
$\Gamma_T(\Gamma_S(\sigma))=\Gamma_{S\cup T}(\sigma)$, expressed 
functorially we have $\Gamma_T\circ\Gamma_S=\Gamma_{S\cup T}$.
\end{thm}

\pr We have 
\[\Gamma_T\circ\Gamma_S=C\circ\st_T\circ C\circ\st_S=
C\circ\st_T\circ\st_S=C\circ\st_{S\cup T}=\Gamma_{S\cup T},\]
where the first and the fourth equalities follow from Definition~\ref{df:go},
the second equality follows from Lemma~\ref{lm:1}, and the third equality 
follows from Proposition~\ref{prop:stst}.
\end{proof} 


\section{Immediate snapshot complexes} \label{sect:3}
\subsection{Round counters} 

\nin
Our main objects of study, the immediate snapshot complexes, are
indexed by finite tuples of nonnegative integers. 

\begin{df}
Given a function $\tr:\zz_+\rightarrow\zz_+\cup\{\bot\}$, we consider
the set
\[\supp\tr:=\{i\in\zz_+\,|\,\tr(i)\neq\bot\}.\] 
This set is called the {\bf support set} of $\tr$.

\nin A {\bf round counter} is a function
$\tr:\zz_+\rightarrow\zz_+\cup\{\bot\}$ with a~finite support set.
\end{df}

Obviously, a~round counter can be thought of as an infinite sequence
$\tr=(\tr(0),\tr(1),\dots)$, where, for all $i\in\zz_+$, either
$\tr(i)$ is a nonnegative integer, or $\tr(i)=\bot$, such that only
finitely many entries of $\tr$ are nonnegative integers. We shall
frequently use a~short-hand notation $\tr=(r_0,\dots,r_n)$ to denote
the round counter given by
\[\tr(i)=\begin{cases}
r_i, & \text{ for } 0\leq i\leq n;\\
\bot,& \text{ for } i>n.
\end{cases}\]

Operationally, the round counter models the number of rounds taken by
each process in the execution of the corresponding distributed
protocol: $r_k$ counts the number of times process $k$ becomes active, where
$r_k=\bot$ means that the process $k$ does not participate in the computation. 

\begin{df}
Given a round counter $\tr$, the number $\sum_{i\in\supp\tr}\tr(i)$ is
called the {\bf cardinality} of $\tr$, and is denoted $|\tr|$. 
The sets
\[\act\tr:=\{i\in\supp\tr\,|\,\tr(i)\geq 1\}
\text{ and } \pass\tr:=\{i\in\supp\tr\,|\,\tr(i)=0\}\] 
are called the {\bf active} and the {\bf passive} sets of~$\tr$.
\end{df}

Assume now we are given a~\pnt $\tr$, and let
  $\varphi:\supp\tr\to[|\supp\tr|-1]$ denote the unique
  order-preserving bijection. The \pnt $c(\tr)$ is defined by
\[c(\tr)(i):=
\begin{cases}
\tr(\varphi^{-1}(i)), &\text{ for } 0\leq i\leq|\supp\tr|-1; \\
\bot,&\text{ for } i\geq|\supp\tr|.
\end{cases}\]
We call $c(\tr)$ the {\it canonical form} of $\tr$. Note that
$\supp c(\tr)=[|\supp\tr|-1]$, $|\act(c(\tr))|=|\act\tr|$,
and $|\pass(c(\tr))|=|\pass(\tr)|$.

Let $\cs_{\zz_+}$ denote the group of bijections $\pi:\zz_+\ra\zz_+$, such that
$\pi(i)\neq\pi(i)$ for only finitely many~$i$. This group acts on
the set of all round counters, namely for $\pi\in\cs_{\zz_+}$, and
a \pnt $\tr$ we set $\pi(\tr)(i):=\tr(\pi(i))$.

If desired by the reader, it is sufficient to think about round counters as finite sequences. However, it is still necessary for some entries to be allowed to take the value $\bot$. This is because the corresponding immediate snapshot complexes have canonical decompositions, whose descriptions are easier if we allow this generality. For the same reason it would not be opportune to sort the entries of $\tr$. On the other hand, permuting the entries of $\tr$ obviously induces simplicial isomorphism. Considering the infinite counters (with only a finite relevant part) and considering the action by the group $\cs_{\zz_+}$ provides a~convenient formal framework to deal with all the cases at once.

\subsection{Combinatorial definition} 


\noindent
We now describe our main objects of study. Our definition proceeds as
follows. We first describe the set of labels of the simplices,
distinguishing the set of labels of the vertices. Then for each
simplex we describe its set of vertices. Clearly, several conditions
will have to be checked later. We have to see that any subset of a~set
of vertices of a~simplex corresponds to some simplex, and crucially,
that any two simplices with the same set of vertices must coincide.

\begin{df}\label{df:ptr}
Assume $\tr$ is a~round counter. We define an abstract simplicial
complex $P(\tr)$, called the {\bf immediate snapshot complex}
associated to the round counter~$\tr$, as follows. The vertices of
$P(\tr)$ are indexed by all witness structures $\sigma=(\{p\},\ab
G,\ab \{\trc(q)\}_{q\in\{p\}\cup G})$, satisfying these three
conditions:
\begin{enumerate}
\item $\{p\}\cup G=\supp\tr$;
\item $|\trc(p)|=\tr(p)+1$;
\item $|\trc(q)|\leq\tr(q)+1$, for all $q\in G$.
\end{enumerate}
\nin We say that such a vertex has {\bf color} $p$. In general, the
simplices of $P(\tr)$ are indexed by all witness structures
$\sigma=(A,\ab G,\ab \{\trc(q)\}_{q\in A\cup G})$, satisfying:
\begin{enumerate}
\item $A\cup G=\supp\tr$;
\item $|\trc(q)|=\tr(q)+1$, for all $q\in A$;
\item $|\trc(q)|\leq\tr(q)+1$, for all $q\in G$.
\end{enumerate}

\nin The empty witness structure $((\es,\supp\tr))$ indexes the empty
simplex of $P(\tr)$. When convenient, we identify the simplices of
$P(\tr)$ with the witness structures which index them.  

Let $\sigma$ be a~non-empty witness structure satisfying the
conditions above. The set of vertices $V(\sigma)$ of the simplex $\sigma$
is given by $\{\Gamma_{A(\sigma)\setminus\{p\}}(\sigma)\,|\,p\in A(\sigma)\}$.
\end{df}

Note, that for an arbitrary witness structure $\sigma$ and
an~arbitrary $p\in A(\sigma)$ we have the equality
$A(\Gamma_{A(\sigma)\sm\{p\}}(\sigma))=\{p\}$. Hence we have
\begin{equation}\label{eq:asi}
A(\sigma)=\{A(v)\,|\,v\in V(\sigma)\},
\end{equation}
in particular, the set of vertices of $\sigma$ uniquely determines
$A(\sigma)$.

\begin{figure}[hbt]
\centering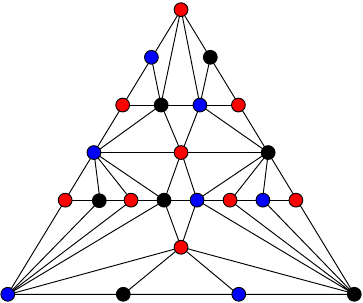
\caption{The immediate snapshot complex $P(2,1,1)$.}
\label{fig:f211}
\end{figure}

Assume $\tr$ is a round counter, such that $\tr(i)=\bot$ for all
$i\geq n+1$.  In line with our short-hand notation for the round
counters, and in addition skipping a pair of brackets, we shall use
an~alternative notation $P(\tr(0),\dots,\tr(n))$ instead of
$P(\tr)$. An~example of an immediate snapshot complex $P(0,1,1)$ is
shown on Figure~\ref{fig:f011}, and a~more sophisticated example
$P(2,1,1)$ is shown on Figure~\ref{fig:f211}.
 
\subsection{The Reconstruction Lemma.} 

\noindent
From the point of view of distributed computing, the vertices of $P(\tr)$
should be thought of as {\it local views} of specific processors. 
In this intuitive picture, the next Reconstruction Lemma~\ref{lm:rec} 
says that any set of local views corresponds to at most one global view.

\begin{lm}\label{lm:rec} {\rm (Reconstruction Lemma).}

\noindent
Assume $\sigma$ and $\tau$ are witness structures of dimension~$d$,
such that the corresponding $d$-simplices of $P(\tr)$ have the same
set of vertices, then we must have $\sigma=\tau$.
\end{lm}
\pr 
Assume the statement of lemma is not satisfied, and pick a~pair
of $d$-dimensional simplices $\sigma\neq\tau$, such that $V(\sigma)=V(\tau)$,
and $d$ is minimal possible. Obviously, we must have $d\geq 1$.

By \eqref{eq:asi}, we have $A(\sigma)=A(\tau)$. Since
$\supp\sigma=\supp\tau=\supp\tr$, we also have
$G(\sigma)=G(\tau)$. For brevity, we set $\Sigma:=A(\sigma)$, and for
each $p\in\Sigma$, we set
$v_p:=\Gamma_{\Sigma\sm\{p\}}(\sigma)=\Gamma_{\Sigma\sm\{p\}}(\tau)$.
For each $p\in\Sigma$, it follows from the definition of the ghosting 
operation that $|\trc(p,\sigma)|=|\trc(p,v_p)|$.
This implies that the $\Sigma$-tuples $(|\trc(p,\sigma)|)_{p\in\Sigma}$ and
$(|\trc(p,\tau)|)_{p\in\Sigma}$ are equal.

Now pick an arbitrary $p\in\Sigma$. For every $q\in\Sigma$, such that $q\neq p$,
we have 
\[\Gamma_{(\Sigma\sm\{p\})\sm\{q\}}(\Gamma_p(\sigma))=\Gamma_{\Sigma\sm\{q\}}(\sigma)=
v_q=\Gamma_{\Sigma\sm\{q\}}(\tau)=\Gamma_{(\Sigma\sm\{p\})\sm\{q\}}(\Gamma_p(\tau)),\]
hence $V(\Gamma_p(\sigma))=V(\Gamma_p(\tau))$. By the minimality of
$\sigma$, this implies $\Gamma_p(\sigma)=\Gamma_p(\tau)$.

Let $\sigma=((W_0,G_0),\dots,(W_t,G_t))$. Assume there exists $0\leq
k\leq t$, and $p,q\in\Sigma$, $p\neq q$, such that
$\last(p,\sigma)=\last(q,\sigma)$. Then, we have
\[\Gamma_p(\sigma)=\begin{array}{|c|c|c|c|c|c|c|}
\hline
W_0 & \dots & W_{k-1} & W_k\sm\{p\}  & W_{k+1} & \dots & W_t \\ \hline
G_0 & \dots & G_{k-1} & G_k\cup\{p\} & G_{k+1} & \dots & G_t \\ 
\hline
\end{array},\]
since $\Gamma_p(\sigma)=\Gamma_p(\tau)$, but $\sigma\neq\tau$, we get
\begin{equation}\label{eq:tau1}
\tau=\begin{array}{|c|c|c|c|c|c|c|c|}
\hline
W_0 & \dots & W_{k-1} & p   & W_k\sm\{p\} & W_{k+1} & \dots & W_t \\ \hline
G_0 & \dots & G_{k-1} & A_p & B_p         & G_{k+1} & \dots & G_t \\ 
\hline
\end{array},\end{equation}
for some $A_p$, $B_p$ such that $A_p\cup B_p=G_k$. Repeating the same argument
with $q$ instead of $p$ we get
\begin{equation}\label{eq:tau2}
\tau=\begin{array}{|c|c|c|c|c|c|c|c|}
\hline
W_0 & \dots & W_{k-1} & q   & W_k\sm\{q\} & W_{k+1} & \dots & W_t \\ \hline
G_0 & \dots & G_{k-1} & A_q & B_q         & G_{k+1} & \dots & G_t \\ 
\hline
\end{array},\end{equation}
for some $A_q$, $B_q$ such that $A_q\cup B_q=G_k$. The
equations~\eqref{eq:tau1} and~\eqref{eq:tau2} contradict each
other. It is thus safe to assume that for every $0\leq k\leq t$, there exists
at most one $p\in\Sigma$, such that $\last(p,\sigma)=k$.

Set $F:=\{p\in\Sigma\,|\,|\trc(p,\sigma)|=|\trc(p,\Gamma_p(\sigma))|\}$.  
Note that \[F=\{p\in\Sigma\,|\,|\trc(p,\tau)|=|\trc(p,\Gamma_p(\tau))|\}.\]  
Using Remark~\ref{rem:tr}, the previous observation $|\trc(p,\sigma)|\leq
|\trc(p,\Gamma_p(\sigma))|$ can be strengthened as follows: we know that
$F=\Sigma\sm\{l\}$, for some $l\in\Sigma$. Specifically, $W_t=\{l\}$,
and the last pair of sets in $\tau$ is also $(\{l\},H)$, for some
$H\subseteq G(\tau)$.

Pick $p\in F$ such that $\last(p)=\max_{q\in F}\last(q)$. Assume 
\[\Gamma_p(\sigma)=\begin{array}{|c|c|c|c|c|c|c|}
\hline
W_0 & \dots & W_{k-1} & W_k          & W_{k+1} & \dots & W_t \\ \hline
G_0 & \dots & G_{k-1} & G_k\cup\{p\} & G_{k+1} & \dots & G_t \\ 
\hline
\end{array}.\]
We observe, that $p$ was chosen so that $(W_k\cup\dots\cup W_t)\cap F=\es$. 
We can easily describe the set $\Lambda$ of all $d$-simplices $\gamma$, for which 
$p\in\supp\gamma$ and $\Gamma_p(\gamma)=\Gamma_p(\sigma)$. Set
\[\gamma^p:=\begin{array}{|c|c|c|c|c|c|c|}
\hline
W_0 & \dots & W_{k-1} & W_k\cup\{p\} & W_{k+1} & \dots & W_t \\ \hline
G_0 & \dots & G_{k-1} & G_k          & G_{k+1} & \dots & G_t \\ 
\hline
\end{array},\]
and
\[\gamma_{A,B}:=\begin{array}{|c|c|c|c|c|c|c|c|}
\hline
W_0 & \dots & W_{k-1} & p & W_k & W_{k+1} & \dots & W_t \\ \hline
G_0 & \dots & G_{k-1} & A &  B  & G_{k+1} & \dots & G_t \\ 
\hline
\end{array},\]
where $A\cup B=G_k$. Then $\Lambda=\{\gamma_{A,B}\,|\,A\cup
B=G_k\}\cup\{\gamma^p\}$.  Clearly, $\sigma,\tau\in\Lambda$. We shall
show that $\Gamma_l(\sigma)\neq\Gamma_l(\tau)$.

Assume $A\cup B=G_k$, and pick $\alpha\in W_k$. Then 
\[|\trc(\alpha,\Gamma_l(\gamma^p))|=\sum_{i=0}^{k-1}\chi(\alpha,R_i)+1\neq
\sum_{i=0}^{k-1}\chi(\alpha,R_i)=|\trc(\alpha,\Gamma_l(\gamma_{A,B}))|,\]
hence $\Gamma_l(\gamma^p)\neq\Gamma_l(\gamma_{A,B})$.

Assume now we have further sets $A'$ and $B'$, such that $A'\cup
B'=G_k$, $A\neq A'$.  Without loss of generality, we can assume that
$A\not\subseteq A'$.  Pick now $\alpha\in A\sm A'$. Then
\[|\trc(\alpha,\Gamma_l(\gamma_{A,B}))|=\sum_{i=0}^{k-1}\chi(\alpha,R_i)+1\neq
\sum_{i=0}^{k-1}\chi(\alpha,R_i)=|\trc(\alpha,\Gamma_l(\gamma_{A',B'}))|,\]
hence $\Gamma_l(\gamma_{A,B})\neq\Gamma_l(\gamma_{A',B'})$.

We have thus proved that $\Gamma_l(\sigma)\neq\Gamma_l(\tau)$,
contradicting the choice of $\sigma$ and~$\tau$.  \end{proof}

\subsection{Immediate snapshot complexes are well-defined} 

\noindent
We are now in a~position to check that the Definition~\ref{df:ptr} yields
a~well-defined simplicial complex, and to show that the ghosting
operation provides the right combinatorial language to describe
boundaries in~$P(\tr)$.

\begin{prop}
\label{prop:b}
Assume $\tr$ is the round counter. 
\begin{itemize}
\item[(1)] The associated immediate snapshot complex $P(\tr)$ is
  a~well-defined simplicial complex. In this complex the dimension of
  the simplex indexed by $\sigma$ is equal to $\dim\sigma$,
\item[(2)] Assume $\sigma$ and $\tau$ are simplices of
$P(\tr)$. Then $\tau\subseteq\sigma$ if and only if there exists 
$S\subseteq A(\sigma)$, such that $\tau=\Gamma_S(\sigma)$.
\end{itemize}
\end{prop}

\pr Assume that the witness structure $\sigma$ indexes a~simplex of
$P(\tr)$. Set $d:=\dim\sigma$, implying that
$A(\sigma)=\{p_0,\dots,p_d\}$ for $p_0<\dots<p_d$, $p_i\in\zz_+$. For
$0\leq i\leq d$, we set
$v_i:=\Gamma_{A(\sigma)\setminus\{p_i\}}(\sigma)$. We see that the
$d$-dimensional simplex $\sigma$ has $d+1$ vertices, which are all
distinct, since $A(v_i)=p_i$, for $0\leq i\leq d$. Furthermore, it
follows from the Reconstruction Lemma~\ref{lm:rec} that any two
simplices with the same set of vertices are equal.

Assume now that $\tau=\Gamma_S(\sigma)$, for some $S\subseteq
A(\sigma)$. By Proposition~\ref{prop:gall} we have
$A(\tau)=A(\sigma)\sm S$.  It follows from Theorem~\ref{thm:gg}
that for every $p\in A(\tau)$ we have
\begin{equation}\label{eq:gatp}
\Gamma_{A(\tau)\sm\{p\}}(\tau)=\Gamma_{A(\tau)\sm\{p\}}(\Gamma_S(\sigma))=
\Gamma_{A(\tau)\cup S\sm\{p\}}(\sigma)=\Gamma_{A(\sigma)\sm\{p\}}(\sigma),
\end{equation} 
hence the set of vertices of $\tau$ is a~subset of the set of vertices
of~$\sigma$.

On the other hand, pick an arbitrary $W\subseteq V(\sigma)$, for some
witness structure~$\sigma$. By definition, there exists $T\subseteq
A(\sigma)$, such that $W=\{\Gamma_{A(\sigma)\setminus\{p\}}\,|\,p\in
T\}$.  Set $\tau:=\Gamma_{A(\sigma)\setminus T}$. The same computation
as in~\eqref{eq:gatp}, shows that $V(\tau)=W$; that is, for any subset
of the set of the vertices of $\sigma$, there exists a~simplex, which
has this subset as its set of vertices.

Finally, assume $V(\tau)\subseteq V(\sigma)$, for some simplices
$\sigma$ and~$\tau$. Again, the same computation as
in~\eqref{eq:gatp}, shows that
$V(\Gamma_{\supp\sigma\sm\supp\tau}(\sigma))=\supp\tau$, i.e., $\tau$
and $\Gamma_{\supp\sigma\sm\supp\tau}(\sigma)$ have the same set of
vertices.  It follows from the Reconstruction Lemma~\ref{lm:rec} that
$\tau=\Gamma_{\supp\sigma\sm\supp\tau}(\sigma)$, and so both statements (1) and
(2) are proved.  \end{proof}


\section{Some observations on immediate snapshot complexes}
\label{sect:4}

\subsection{Elementary properties and examples} 

\noindent
We start by listing a~few simple but useful properties of the
immediate snapshot complexes~$P(\tr)$.  

\begin{prop}\label{prop:p1} 
For an arbitrary point counter $\tr$, we have 
\begin{equation}\label{eq:cfe}
P(\tr)\simeq P(c(\tr)),
\end{equation} 
where $\simeq$ denotes an isomorphism of simplicial complexes.
\end{prop}
\pr
Consider the map
\[\Phi:((W_0,G_0),\dots,(W_t,G_t))\mapsto
((\varphi(W_0),\varphi(G_0)),\dots,(\varphi(W_t),\varphi(G_t))),\]
where $\varphi$ is the unique order-preserving bijection
$\varphi:\supp\tr\rightarrow[|\supp\tr|-1]$. This gives a~bijection
between simplices of $P(\tr)$ and simplices of $P(c(\tr))$. Since
$\varphi$ is just a~renaming bijection, we conclude that $\Phi$ is
a~simplicial isomorphism. \end{proof}

\vspace{5pt} 

\nin In particular, if \pnts $\tr$ and $\tq$ have the same canonical
form, then the corresponding immediate snapshot complexes are
isomorphic. In other words, the $\bot$ entries do not matter for the
simplicial structure. This can be generalized as follows.

\begin{prop}\label{prop:p2} 
For any \pnt $\tr$, and any permutation $\pi\in\csn$, the simplicial
complex $P(\pi(\tr))$ is isomorphic to the simplicial
complex~$P(\tr)$.
\end{prop}
\pr Consider the map
\[\Phi:((W_0,G_0),\dots,(W_t,G_t))\mapsto
((\pi(W_0),\pi(G_0)),\dots,(\pi(W_t),\pi(G_t))).\] This map is a
simplicial isomorphism for the same reasons as in the proof of
Proposition~\ref{prop:p1}. \end{proof} \vskip5pt

Let us now look at special round counters. If $\tr=(r)$, then the
simplicial complex $P(\tr)$ is just a~point indexed by the witness
structure $(\underbrace{(0,\es),\dots,(0,\es)}_{r+1})$. Recall, that
the empty simplex of $P(r)$ is indexed by the witness structure
$((\es,0))$.

\begin{prop}\label{prop:p3}
The immediate snapshot complex $P(\underbrace{0,\dots,0}_{n+1})$ is
isomorphic as a~simplicial complex to the $n$-simplex
$\Delta^{n}$. More generally, if $\tr$ is a round counter such that
$r(i)\in\{\bot,0\}$, for all $i\in\zz_+$, the simplicial complex
$P(\tr)$ is isomorphic with $\Delta^{\supp\tr}$.
\end{prop}
\pr The simplices of $P(\tr)$ are indexed by all $((A,B))$ such that
$A\cap B=\es$ and $A\cup B=[n]$. The simplicial isomorphism between
$P(\tr)$ and $\da^n$ is given by $((A,B))\mapsto A$. The second
statement follows from Proposition~\ref{prop:p1}. \end{proof}

\begin{prop}\label{prop:p4}
Assume $\tr=(r(0),\dots,r(n))$ and $\tr(n)=0$. Let $\bar q$ denote the
truncated round counter $(r(0),\dots,r(n-1))$. Consider a~cone over
$P(\bar q)$, which we denote $P(\bar q)*\{a\}$, where $a$ is the apex
of the cone. Then we have
\begin{equation}\label{eq:ptr} 
P(\tr)\simeq P(\bar q)*\{a\}.
\end{equation} 
\end{prop}
\pr Let $\sigma=((W_0,G_0),\dots,(W_t,G_t))$ be a simplex of $P(\tr)$
and consider the map
\[\Phi:\sigma\mapsto
\begin{cases}
((W_0\setminus\{n\},G_0),\dots,(W_t,G_t))*\{a\},&\textrm{if } n\in W_0;\\
((W_0,G_0\setminus\{n\}),\dots,(W_t,G_t)),&\textrm{if } n\in G_0.
\end{cases}\]
Since $W_0\cup G_0=[n]$, and $W_0\cap G_0=\emptyset$, we either have
$n\in W_0$ or $n\in G_0$. If $n\in G_0$, then $n\notin
W_0\cup\dots\cup W_t\cup\ab G_1\cup\dots\cup G_t$. If $n\in W_0$, then
$n\notin G_0$. Furthermore, since $\tr(n)=0$, we have
$|\trc(n,\sigma)|\leq 1$, hence $\trc(n,\sigma)=\{0\}$, and $n\notin
W_1\cup\dots\cup W_t\cup\ab G_0\cup\dots\cup G_t$. In any case, $\Phi$
is well-defined. Its inverse is also clear, so it is a~bijection between 
simplices of $P(\tr)$ and $P(\bar q)*\{a\}$.

Under this bijection, the vertex $((n,[n-1]))$ of $P(\tr)$ corresponds
to the apex~$a$. The map $\Phi$ is simplicial, since ghosting other
elements than $n$ will be for both complexes; while ghosting the
element $n$ will simply move it from $W_0$ to $G_0$ in a~simplex from
$P(\tr)$, which corresponds to the deletion of the apex $a$ in
a~simplex from $P(\bar q)*\{a\}$.  \end{proof}

\vskip5pt

\nin Clearly, the applications of Proposition~\ref{prop:p4} can be
iterated, until each $0$ entry in $\tr$ is replaced with a~cone
construction.

The Propositions~\ref{prop:p1}, \ref{prop:p2}, \ref{prop:p3}, and
\ref{prop:p4}, can intuitively be summarized as telling us that if we
are interested in understanding the simplicial structure of the
complex $P(\tr)$, we may ignore the entries $\bot$ and $0$, and
permute the remaining entries as we see fit.

\subsection{The purity of the immediate snapshot complexes} 

\nin Assume $\sigma=((W_0,G_0),\dots,(W_t,G_t))$ is a~witness
structure which indexes a~simplex of $P(\tr)$. Clearly, we have
$|A(\sigma)|\leq|\supp\tr|$, hence $\dim\sigma\leq|\supp\tr|-1$. It
turns out that every simplex is contained in a~simplex of dimension
$|\supp\tr|-1$, which is the same as to say that immediate snapshot
complexes are always pure (that is all maximal simplices have the same
dimension).

\begin{prop}\label{prop:pure}
The simplicial complex $P(\tr)$ is pure of dimension $|\supp\tr|-1$.
\end{prop}
\pr Assume $\sigma=((W_0,G_0),\dots,(W_t,G_t))$ is a~witness structure
which indexes a~simplex of $P(\tr)$. For each $p\in G(\sigma)$ we set
$m(p):=r(p)+1-|\trc(p,\sigma)|$. By construction, we have $m(p)\geq 0$.
Set furthermore $q:=\max_{p\in G(\sigma)} m(p)$,
\[V_i:=\{p\in G(\sigma)\,|\,m(p)\geq i\}, \textrm{ for } i=1,\dots,q,\] and
\[\tilde\sigma:=(W_0\cup G_0,W_1\cup G_1,\dots,W_t\cup G_t,V_1,\dots,V_q).\]
We see that $\tilde\sigma$ is a~witness structure: the condition (P1)
says that $V_i\subseteq W_0\cup G_0$, which is clear, the conditions
(P2) and (P3) are immediate, and condition (W) says that $V_i\neq 0$,
which is also clear. Furthermore, we have
$\supp\tilde\sigma=\supp\sigma$, $G(\tilde\sigma)=\es$, and
$A(\tilde\sigma)=\supp\sigma=A(\sigma)\cup G(\sigma)$. For all
$\sigma\in A(\sigma)$ we have
\[|\trc(p,\tilde\sigma)|=|\trc(p,\sigma)|=r(p)+1,\] while for all
$\sigma\in G(\sigma)$ we have
\[|\trc(p,\tilde\sigma)|=|\trc(p,\sigma)|+m(p)=r(p)+1.\] 
We conclude that $\tilde\sigma$ indexes a~simplex of
$P(\tr)$. Clearly, $\dim\tilde\sigma=|\supp\sigma|-1$. Finally, we
have $\Gamma(\tilde\sigma,G(\sigma))=\sigma$, so, by
Proposition~\ref{prop:b}(2), $\tilde\sigma\subseteq\sigma$ and hence
$P(\tr)$ is pure of dimension~$|\supp\tr|-1$. \end{proof}

\subsection{Immediate snapshot complexes of dimension $1$} 

\nin It follows from the above, that $\dim P(\tr)=0$ if and only if
$|\supp\tr|=1$, in which case the simplicial complex $P(\tr)$ is
a~point. Assume now $\dim P(\tr)=1$. In this case, we have
$|\supp\tr|=2$. By \eqref{eq:cfe}, up to the simplicial isomorphism,
we can assume that $\tr=(m,n)$, $m,n\geq 0$.

For brevity of notations, when talking about edges of $P(m,n)$, we
shall skip $W_0=[1]$, and index the edges by the tuples
$(W_1,\dots,W_t)$ of subsets $W_i\subseteq[1]$, such that
$\sum_{i=1}^t\chi(0,W_i)=m$ and $\sum_{i=1}^t\chi(1,W_i)=n$. We shall
make no abbreviations when indexing the vertices.

\begin{prop}\label{prop:pmn}
For any integers $m,n\geq 0$, the simplicial complex $P(m,n)$ is
a~subdivided interval, whose endpoints are indexed by
\[v^0_{m,n}:=((0,1),\ab\underbrace{(0,\es),\dots,\ab(0,\es)}_m),\textrm{ and } 
v^1_{m,n}:=((1,0),\ab\underbrace{(1,\es),\dots,\ab(1,\es)}_n).\]
\end{prop}
\pr To start with, we know that the simplicial complex $P(m,n)$ is
a~pure $1$-dimensional complex, and that without loss of generality,
we can assume $m\geq n$. The simplicial complex $P(m,0)$ is just a~$1$-simplex,
indexed by $(\underbrace{0,\dots,0}_m)$, and the claim of proposition is 
obviously true in this case. Our proof now makes use of induction on $m+n$.

When proving the induction step, we are free to confine ourselves to
the case $m\geq n\geq 1$. For the sets $S=\{0\}$, $\{1\}$, and
$\{0,1\}$, we let $A_S$ denote the pure $1$-dimensional subcomplex of
$P(m,n)$, obtained by taking the union of all $1$-simplices of the
form $(S,S_1,\dots,S_t)$. For brevity, we shall simply write $A_0$, $A_1$,
and $A_{01}$. Obviously, each $1$-simplex of $P(m,n)$ belongs precisely
to one of these three sets, so we have $A_0\cup A_1\cup A_{01}=P(m,n)$,
and we shall now see how the three subcomplexes fit together.

It is easy to see, either directly, or as a~special case of
Proposition~\ref{prop:strata2}, that we have simplicial isomorphisms
$A_0\cong P(m-1,n)$, $A_1\cong P(m,n-1)$, and $A_{01}\cong
P(m-1,n-1)$. Consider now two special vertices of $P(m,n)$
\[w_0=(([1],\es),(0,1),\underbrace{(0,\es),\dots,\ab(0,\es)}_{m-1}),
\textrm{ and }
w_1=(([1],\es),(1,0),\underbrace{(1,\es),\dots,\ab(1,\es)}_{n-1}).\]

By the induction assumption, each one of the subcomplexes $A_0$, $A_1$,
and $A_{01}$, is a~subdivided interval, and by the same assumption,
combined with the simplicial isomorphism from
Proposition~\ref{prop:strata2}, we know what the endpoints are.
Namely, $A_0$ has endpoints $v^0_{m,n}$ and $w_1$, $A_1$ has endpoints
$v^1_{m,n}$ and $w_0$, and $A_{01}$ has endpoints $w_0$ and
$w_1$. Obviously, that means that these three subcomplexes piece
together to form a~new subdivided interval with endpoints $v^0_{m,n}$
and $v^1_{m,n}$.  \end{proof}

\vskip5pt

Let $f(m,n)$ denote the number of $1$-simplices in $P(m,n)$. By
Proposition~\ref{prop:pmn} this number completely describes the
complex $P(m,n)$. We do not have a~closed formula for these numbers,
however, we can explicitly describe the corresponding generating
function.

\begin{prop}
The numbers $f(m,n)$ satisfy the recursive relation
\begin{equation}\label{eq:fmn}
f(m,n)=f(m,n-1)+f(m-1,n)+f(m-1,n-1),\quad\forall m,n\geq 1,
\end{equation}
with the boundary conditions $f(m,0)=f(0,m)=1$. The corresponding
generating function
\[F(x,y)=\sum_{m,n=0}^\infty f(m,n)x^m y^n\]
is given by the following explicit formula:
\begin{equation}\label{eq:fxy}
F(x,y)=\dfrac{1}{1-x-y-xy}.
\end{equation}	
\end{prop}
\pr The fact that $f(m,0)=f(0,m)=1$, as well as that $f(m,n)=f(n,m)$,
are both immediate. Assume now that $m,n\geq 1$. The number of edges
of $P(m,n)$ for which $W_1=\{0\}$ is $f(m-1,n)$, the number of edges
of $P(m,n)$ for which $W_1=[1]$ is $f(m-1,n-1)$, finally, the number
of edges of $P(m,n)$ for which $W_1=\{1\}$ is $f(m,n-1)$. Summing up we 
get the recursive formula~\eqref{eq:fmn}.

Multiply~\eqref{eq:fmn} with $x^m y^n$ and sum over all $m,n\geq 1$.
We get
\begin{equation}\label{eq:fmns}
\sum_{m,n\geq 1}f(m,n)x^m y^n=\sum_{m\geq 1,n\geq 0}f(m,n)x^m y^{n+1}+
\sum_{m\geq 0,n\geq 1}f(m,n)x^{m+1}y^n+\sum_{m,n\geq 0}f(m,n)x^{m+1}y^{n+1}.
\end{equation}
On the left hand side we have
\[\sum_{m,n\geq 1}f(m,n)x^m y^n=F(m,n)-1-\sum_{m\geq 1}x^m-\sum_{n\geq 1}y^n=
F(x,y)-\frac{1}{1-x}-\frac{1}{1-y}+1.\] 
On the right hand side we have
\[\sum_{m\geq 1,n\geq 0}f(m,n)x^m y^{n+1}=y\cdot\sum_{m\geq 1,n\geq 0}f(m,n)x^m y^n=
y\left(F(x,y)-\sum_{n\geq 0}y^n\right)=y\left(F(x,y)-\frac{1}{1-y}\right).\] 
Transforming the other terms on the right hand side of \eqref{eq:fmns}
in a similar way, we get
\[F(x,y)-\frac{1}{1-x}-\frac{1}{1-y}+1=xF(x,y)-\frac{x}{1-x}+yF(x,y)-
\frac{y}{1-y}+xyF(x,y),\] 
which simplifies to $F(x,y)(1-x-y-xy)=1$ yielding the formula~\eqref{eq:fxy}.  
\end{proof}

\subsection{Number of simplices of maximal dimension in an immediate 
snapshot complex.} 

\nin For arbitrary nonnegative integers $m_0,\dots,m_n$ we let
$f(m_0,\dots,m_n)$ denote the number of top-dimensional simplices in
$P(m_0,\dots,m_n)$. Note, that the top-dimensional simplices of
$P(m_0,\dots,m_n)$ are indexed by sequences $(W_1,\dots,W_t)$ of
non-empty subsets $W_i\subseteq[n]$, such that
$\sum_{i=1}^t\chi(p,W_i)=m_p$, for all $p\in[n]$.
\begin{prop}
We have 
\[f(m_0,\dots,m_{n-1},0)=f(m_0,\dots,m_{n-1}),\] 
and also
\[f(m_0,\dots,m_n)=f(m_{\pi(0)},\dots,m_{\pi(n)}),\] for any
$\pi\in\cs_{[n]}$. 

In general, consider a round counter $\tr=(m_0,\dots,m_n)$, then we have
\begin{equation}\label{eq:fmn2}
f(m_0,\dots,m_n)=\sum_{\es\neq S\subseteq\act\tr}f(m_0^S,\dots,m_n^S),
\end{equation}
where 
\[m_k^S=\begin{cases}
m_k-1, & \textrm{ if }k\in S;\\
m_k, & \textrm{ if }k\notin S.
\end{cases}\]
The corresponding generating function in $n+1$ variables is
\[F(x_0,\dots,x_n)=\sum_{m_0,\dots,m_n=0}^\infty f(m_0,\dots,m_n)x_0^{m_0}\dots x_n^{m_n}.\]
It is given by the following explicit formula:
\begin{equation}\label{eq:fxn}
F(x_0,\dots,x_n)=1/\left(1-\sum_{\es\neq S\subseteq\act\tr}\prod_{j\in S} x_j\right).
\end{equation}	
\end{prop}
\pr The first two equalities are immediate. To prove the
equality~\eqref{eq:fmn2} simply sum over the top-dimensional simplices
grouping them according to the subset~$W_1$.  The formula
\eqref{eq:fxn} can either be derived same way as we derived the
formula \eqref{eq:fxy}, or by a~term-by-term calculation of the
product 
\[F(x_0,\dots,x_n)\cdot\ab\left(1-\sum_{\es\neq
  S\subseteq\act\tr}\prod_{j\in S} x_j\right)\] using the recursive
formula~\eqref{eq:fmn2}.  \end{proof}

\subsection{Standard chromatic subdivision as immediate snapshot complex} 
\label{ssect:scd}

\nin The {\it standard chromatic subdivision} of an $n$-simplex,
denoted $\chi(\da^n)$, is a~prominent and much studied structure in
distributed computing. We refer to \cite{HKR, HS} for distributed
computing background, and to \cite{subd,view} for the analysis of its
simplicial structure, where, in particular, the following
combinatorial description of $\chi(\da^n)$ has been given.

\begin{df}  \label{df:chi}
Let $n$ be a natural number. The simplicial complex $\chi(\da^n)$ is
defined as follows.
\begin{itemize}
\item The vertices of $\chi(\da^n)$ are indexed by all pairs $(p,V)$, such that 
$V\subseteq[n]$, and $p\in V$.
\item The simplices of $\chi(\Delta^n)$ are indexed by pairs of tuples of
non-empty sets $((B_1,\dots,\ab B_t)\ab(C_1,\dots,C_t))$, such that $B_i$'s
are disjoint subsets of $[n]$, and $C_i\subseteq B_i$ for all~$1\leq i\leq t$.
\end{itemize}
Given a~simplex $\tau=((B_1,\dots,B_t),(C_1,\dots,C_t))$, its vertices
are indexed by all pairs $(c,B)$, where $c\in C_i$, and $B=B_i$, for
some $1\leq i\leq t$.
\end{df}

\nin In particular, the dimension of the simplex $\tau$ indexed by
$((B_1,\dots,B_t)\ab(C_1,\dots,C_t))$ is equal to
$|C_1|+\dots+|C_t|-1$.  To describe the boundary relations in
$\chi(\da^n)$ pick $p\in C_1\cup\dots\cup C_t$, and let $\sigma_p$ be
the simplex obtained from $\tau$ by deleting~$p$. Assume $p\in C_k$.
If $|C_k|\geq 2$, then we have
\begin{equation}\label{eq:si1}
\sigma_p=((B_1,\dots,B_t),(C_1,\dots,C_k\setminus\{p\},\dots,C_t)).
\end{equation}
Otherwise, we have $|C_k|=1$, i.e., $C_k=\{p\}$. If $k<t$, then we have
\begin{equation}\label{eq:si2}
\sigma_p=((B_1,\dots,B_{k-1},B_k\cup B_{k+1},B_{k+2},\dots,B_t),
(C_1,\dots,C_{k-1},C_{k+1},C_{k+2},\dots,C_t)),
\end{equation}
else $k=t$, and we have
\begin{equation}\label{eq:si3}
\sigma_p=((B_1,\dots,B_{t-1}),(C_1,\dots,C_{t-1})).
\end{equation}

For brevity, we set $P_n:=P(\underbrace{1,\dots,1}_{n+1})$.

\begin{prop}
The immediate snapshot complex $P_n$ and the standard chromatic
subdivision of an $n$-simplex $\chi(\da^n)$ are isomorphic as
simplicial complexes. Explicitly, the isomorphism can be given by
\begin{equation}\label{eq:bc}
\Phi:((B_1,\dots,B_t)(C_1,\dots,C_t))\mapsto
\begin{array}{|c|c|c|c|c|}
\hline
W_0 & C_1        & C_2        & \dots & C_t \\ \hline
[n]\sm W_0 & B_1\sm C_1 & B_2\sm C_2 & \dots & B_t\sm C_t \\ 
\hline
\end{array},
\end{equation}
where $W_0=B_1\cup\dots\cup B_t$.
\end{prop}
\pr Let $\tau=((B_1,\dots,B_t)(C_1,\dots,C_t))$ be a simplex of
$\chi(\da^n)$.  We can verify that $\Phi(\tau)$ is a~well-defined
witness structure: (P1) is true since $W_0=B_1\cup\dots\cup B_t$, (P2)
and (P3) are true since the sets $B_i$ are disjoint, while (W) is
true, since the sets $C_i$ are non-empty. We have
$\supp(\Phi(\tau))=[n]$, and $A(\Phi(\tau))=C_1\cup\dots\cup
C_t$. Furthermore, to see that the witness structure $\Phi(\tau)$
indexes a simplex of $P_n$, we notice that $|\trc(p)\leq 2|$, for all
$p\in[n]$, follows from the disjointness of the sets $B_i$, and that
$|\trc(p)|=2$ if and only if $p\in C_1\cup\dots\cup C_t$. We have
\[\dim(\tau)=|C_1|+\dots+|C_t|-1=\dim(\Phi(\tau)).\] 
Finally, the case-by-case comparison of the equations~\eqref{eq:si1},
\eqref{eq:si2}, and~\eqref{eq:si3}, with the rules of the ghosting
operations shows that the map $\Phi$ is simplicial.

Let now $\sigma=((W_0,G_0),\dots,(W_t,G_t))$ be a simplex of $P_n$. Define
\begin{equation}\label{eq:bc2}
\Psi:\sigma\mapsto((W_1\cup G_1,\dots,W_t\cup G_t),(W_1,\dots,W_t)).
\end{equation}
Set $((B_1,\dots,B_t)(C_1,\dots,C_t)):=\Psi(\sigma)$. Clearly,
$C_i\subseteq B_i$, for all $i$, and the sets $C_i$ are non-empty,
since $\sigma$ is a~witness structure. The disjointness of the sets
$B_i$ is immediate consequence of the inequality $|\trc(p)|\leq 2$,
for all $p\in[n]$. It follows that $\Psi(\sigma)$ is a~simplex of
$\chi(\da^n)$.  Obviously, $\Psi$ is an inverse of $\Phi$, hence
$\Phi$ is a~simplicial isomorphism between $\chi(\da^n)$
and~$P_n$. \end{proof}

\vskip5pt

We note the following direct description of the simplicial structure
of $P_n$: simplices of $P_n$ are indexed by all witness structures
$\sigma=((W_0,G_0),\dots,(W_t,G_t))$ satisfying the following three
conditions:
\begin{enumerate} 
	\item [(1)] $W_0\cup G_0=[n]$;
	\item [(2)] $W_0=W_1\cup\dots\cup W_t\cup G_1\cup\dots\cup G_t$;
	\item [(3)] the sets $W_1,\dots,W_t,G_1,\dots,G_t$ are disjoint.
\end{enumerate}

\section{Topology of the immediate snapshot complexes} \label{sect:5}
\subsection{A canonical decomposition of the immediate snapshot complexes} 

\nin We shall now describe how to decompose the immediate snapshot
complex $P(\tr)$ into pieces in a~natural way, which we call the
canonical decompositions. Intuitively, these pieces correspond to the
protocol complexes, for the sets of executions where the first
execution step is fixed.

\begin{df}
Assume $\tr$ is a~round counter. For every subset $S\subseteq\act\tr$,
let $X_S(\tr)$ denote the set of all simplices
$\sigma=((W_0,G_0),\dots,\ab (W_t,G_t))$ of $P(\tr)$, such that one of
the following three conditions is fulfilled:
\begin{itemize}
\item $t=0$; 
\item $S\subseteq G_1$;
\item $W_1\cup G_1=S$.
\end{itemize}
\end{df}

\begin{figure}[hbt]
\centering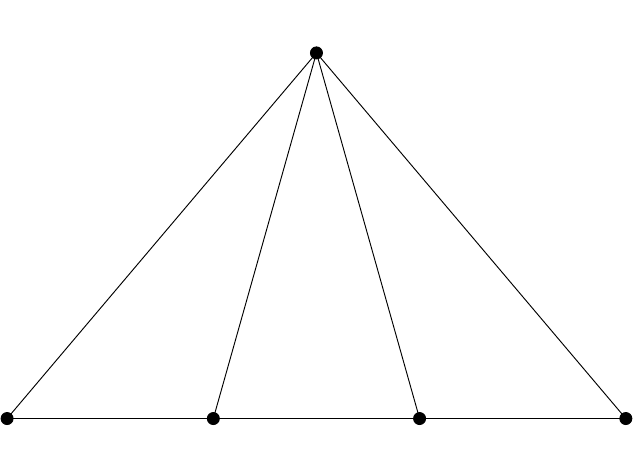
\caption{The immediate snapshot complex $P(0,1,1)$ and its canonical
  decomposition.}
\label{fig:f011}
\end{figure}

\nin In particular, for $S=\es$ the condition $S\subseteq G_1$ is
always satisfied, so $X_\es(\tr)$ is the set of all simplices of
$P(\tr)$.  An example of a canonical decomposition is given on
Figure~\ref{fig:f011}.

\begin{prop}
For every round counter $\tr$, and for every subset
$S\subseteq\act\tr$, the set $X_S(\tr)$ is closed under taking
boundary, hence forms a~simplicial subcomplex of $P(\tr)$.
\end{prop}
\pr Let $\sigma=((W_0,G_0),\dots,(W_t,G_t))$ be a~simplex in
$X_S(\tr)$, and assume $\tau\subset\sigma$. By
Proposition~\ref{prop:b} there exists $T\subseteq A(\sigma)$, such
that $\tau=\Gamma_T(\sigma)$. By Theorem~\ref{thm:gg} it is
enough to consider the case $|T|=1$, so assume $T=\{p\}$, and let
$\tau=((\wti W_0,\wti G_0),\ab\dots,\ab(\wti W_{\tilde t},\wti G_{\tilde
  t}))$.

If $\tilde t=0$, then $\tau\in X_S(\tr)$, and we are done. So assume
$\tilde t\geq 1$, hence also $t\geq 1$. In this case, by definition of
$X_S(\tr)$, we have either $S\subseteq G_1$ or $W_1\cup G_1=S$. On the
other hand, by the definition of $\Gamma_p(\sigma)$, if $\tilde t\geq
1$, then either $W_1\cup G_1\subseteq\wti G_1$ (if all of $W_1$ is
moved to $G_1$) or $W_1\cup G_1=\wti W_1\cup\wti G_1$ and $\wti
G_1\supseteq G_1$ (if only part or none of $W_1$ is moved to $G_1$).

First, if $W_1\cup G_1\subseteq\wti G_1$, then in any case
$S\subseteq\wti G_1$, so $\tau\in X_S(\tr)$, and we are done. Finally,
assume $W_1\cup G_1=\wti W_1\cup\wti G_1$ and $\wti G_1\supseteq
G_1$. If $W_1\cup G_1=S$, then also $\wti W_1\cup \wti G_1=S$, and
$\tau\in X_S(\tr)$. If, instead, $S\subseteq G_1$, then $S\subseteq
\wti G_1$, so again $\tau\in X_S(\tr)$.  \end{proof}

\vskip5pt

\nin We shall abuse notations and use $X_S(\tr)$ to denote this
simplicial complex as well. Next we prove that the subcomplexes
$X_S(\tr)$ can themselves be viewed as immediate snapshot
complexes. To formulate this result we need additional terminology.

\begin{df} Assume $\tr$ is an arbitrary round counter and
$S\subseteq\act\tr$. We let $\tr\dar S$ denote the round counter
  defined by
\[(\tr\dar S)(i)=\begin{cases}
\tr(i), & \text{ if } i\notin S;\\
\tr(i)-1,   & \text{ if } i\in S.
\end{cases}\]
\end{df}

\nin We say that the round counter $\tr\dar S$ is obtained from $\tr$
by the {\it execution} of~$S$. Note that $\supp(\tr\dar S)=\supp\tr$,
$\act(\tr\dar S)=\{i\in\act\tr\,|\, i\notin S, \textrm{ or }\tr(i)\geq
2 \}$, and $\pass(\tr\dar S)=\pass(\tr)\cup\{i\in S\,|\,\tr(i)=1\}$.

\begin{prop}\label{prop:strata2}
Assume $\tr$ is an arbitrary round counter and $S\subseteq\act\tr$,
then there exists a~simplicial isomorphism
\[\gamma_S(\tr):X_S(\tr)\rightarrow P(\tr\dar S).\] 
\end{prop}
\pr Pick an~arbitrary simplex $\sigma=((W_0,G_0),\ab\dots,(W_t,G_t))$
belonging to $X_S(\tr)$. If $t=0$, then we set
$\gamma_S(\sigma):=\sigma$. Note, that since $S\subseteq\act\tr$, we
have $S\subseteq G_0$ in this case. Else, by the construction of
$X_S$, we either have $W_1\cup G_1=S$, or $S\subseteq G_1$. If
$W_1\cup G_1=S$, then set
\[\gamma_S(\sigma):=\begin{array}{|c|c|c|c|}
\hline
W_0\sm G_1  & W_2 & \dots & W_t \\ \hline
G_0\cup G_1 & G_2 & \dots & G_t \\ 
\hline
\end{array},\]
else $S\subseteq G_1$, in which case we set
\[\gamma_S(\sigma):=\begin{array}{|c|c|c|c|c|}
\hline
W_0\sm  S & W_1      & W_2 & \dots & W_t \\ \hline
G_0\cup S & G_1\sm S & G_2 & \dots & G_t \\ 
\hline
\end{array}.\]

Reversely, assume $\tau=((V_0,H_0),\dots,(V_t,H_t))$ is a~simplex of
$P(\tr\dar S)$. Since $\supp\tr=\supp(\tr\dar S)$, we have $S\subseteq
V_0\cup H_0$. If $V_0\cap S\neq\es$, we set
\[\rho_S(\tau):=\begin{array}{|c|c|c|c|c|}
\hline
V_0\cup(H_0\cap S) & V_0\cap S & V_1 & \dots & V_t \\ \hline
H_0\sm(H_0\cap S)  & H_0\cap S & H_1 & \dots & H_t \\ 
\hline
\end{array}.\]
Else we have $S\subseteq H_0$. If $t\geq 1$, we set
\[\rho_S(\tau):=\begin{array}{|c|c|c|c|c|}
\hline
V_0\cup S & V_1       & V_2 & \dots & V_t \\ \hline
H_0\sm  S & H_1\cup S & H_2 & \dots & H_t \\ 
\hline
\end{array},\]
else $t=0$, and we set $\rho_S(\tau):=\tau$.

A~direct case-by-case verification shows that the maps $\gamma_S$ and 
$\rho_S$ are well-defined simplicial maps, which preserve supports,
$A(-)$, $G(-)$, and hence also the dimension. Furthermore, they are 
inverses of each other, hence are simplicial isomorphisms.
\end{proof}

\subsection{Immediate snapshot complexes are pseudomanifolds with boundary} 

\nin In this section we show that immediate snapshot complexes are
pseudomanifolds with boundary.  We start by showing that $P(\tr)$ is
strongly connected.

\begin{df}
Let $K$ be a pure simplicial complex of dimension $n$. Two
$n$-simplices of $K$ are said to be {\bf strongly connected} if there
is a~sequence of $n$-simplices so that each pair of consecutive
simplices has a~common $(n-1)$-dimensional face. The complex $K$ is
said to be {\bf strongly connected} if any two $n$-simplices of $K$
are strongly connected.
\end{df}

Clearly, being strongly connected is an equivalence relation on the
set of all $n$-simplices.

\begin{prop}\label{prop:strc}
For an arbitrary round counter $\tr$, the immediate snapshot complex
$P(\tr)$ is strongly connected.
\end{prop}
\pr Set $n:=|\supp\tr|-1$. Proposition~\ref{prop:pure} says that
$P(\tr)$ is a pure simplicial complex of dimension~$n$. We now use
induction on $|\tr|$. If $|\tr|=0$, or more generally, if
$|\act\tr|\leq 1$, then $P(\tr)$ is just a~single simplex,
so it is trivially strongly connected. 

Assume $|\act\tr|\geq 2$, and consider the canonical decomposition of
$P(\tr)$.  By Proposition~\ref{prop:strata2}, 
the simplicial complex $X_S(\tr)$ is isomorphic to $P(\tr\dar S)$, 
for all $S\subseteq\act\tr$. Since $|\tr\dar S|=|\tr|-|S|<|\tr|$, and
$\supp\tr\dar S=\supp\tr$, we conclude that $X_S(\tr)$ is a~pure
simplicial complex of dimension~$n$, which is strongly connected by
the induction assumption. Thus, any pair of $n$-simplices belonging to
the same subcomplex $X_S(\tr)$ is strongly connected.

Pick now any $p\in\act\tr$, and any $S\subseteq\act\tr$, such that
$p\in S$, $\{p\}\neq S$, and consider any $(n-1)$-simplex 
$\tau=((W_0,G_0),\dots,(W_t,G_t))$, such that $(W_1,G_1)=(S\sm\{p\},\{p\})$. 
Obviously, such $\tau$ exists, and $\tau\in X_S(\tr)\cap X_{p}(\tr)$. 
By induction assumptions for $X_S(\tr)$ and $X_{p}(\tr)$, there exist
$n$-simplices $\sigma_1\in X_S(\tr)$, and $\sigma_2\in
X_{p}(\tr)$, such that $\tau\in\partial\sigma_1$ and
$\tau\in\partial\sigma_2$.  This means, that $\sigma_1$ and $\sigma_2$
are strongly connected.  Since being strongly connected is an
equivalence relation, any two $n$-simplices from $X_S(\tr)$ and
$X_{p}(\tr)$ are strongly connected. This includes the case
$S=\act\tr$, implying that any pair of $n$-simplices in $P(\tr)$ is
strongly connected, so $P(\tr)$ itself is strongly connected. 
\end{proof}

\begin{df}
We say that a~strongly connected pure simplicial complex $K$ is a~{\bf
  pseudomanifold} if each $(n-1)$-simplex of $K$ is a~face of
precisely one or two $n$-simplices of $K$. The $(n-1)$-simplices of
$K$ which are faces of precisely one $n$-simplex of $K$ form
a~(possibly empty) simplicial subcomplex of $K$, called the {\bf
  boundary} of $K$, and denoted $\partial K$.
\end{df}

To describe the boundary subcomplex of $P(\tr)$, we need the following
definition.

\begin{df}\label{df:bv}
Let $\tr$ be an arbitrary round counter, and assume
$V\subset\supp\tr$. We define $B_V(\tr)$ to be the simplicial
subcomplex of $P(\tr)$ consisting of all simplices
$\sigma=((W_0,G_0),\dots,\allowbreak (W_t,G_t))$, satisfying
$V\subseteq G_0$.
\end{df}

\begin{prop}\label{prop:pseudo}
For an arbitrary round counter $\tr$, the simplicial complex $P(\tr)$
is a~pseudomanifold, and the subcomplex $\partial P(\tr)$ consists of
all simplices $\sigma=((W_0,G_0),\dots,(W_t,G_t))$, such that
$G_0\neq\es$.
\end{prop}
\pr By Proposition~\ref{prop:strc} we already know that $P(\tr)$ is
strongly connected. Set again $n:=|\supp\tr|-1$, and let 
$\tau=((W_0,G_0),\dots,(W_t,G_t))$ be an arbitrary $(n-1)$-simplex of $P(\tr)$.
Note that $\codim\tau=|G_0|+\dots+|G_t|$, hence $\codim\tau=1$ implies that
there exist $0\leq k\leq t$, and $p\in\supp\tr$, such that 
\[G_i=\begin{cases}
\{p\},& \textrm{ if } i=k;\\
\es,  & \textrm{ if } i\neq k.
\end{cases}\]
Set $m:=r(p)+1-|\trc(p,\sigma)|$. Consider
\[\sigma_1=(W_0,\dots,W_k-1,W_k\cup\{p\},W_{k+1},\dots,W_t,
\underbrace{p,\dots,p}_m),\]  
and if $k\geq 1$, consider also
\[\sigma_2=(W_0,\dots,W_k-1,p,W_k,\dots,W_t,
\underbrace{p,\dots,p}_m).\] Obviously,
$\Gamma(\sigma_1,p)=\Gamma(\sigma_2,p)=\tau$, so
$\tau\in\partial\sigma_1$ and $\tau\in\partial\sigma_2$. Furthermore,
the definition of the ghosting construction implies that these are the
only options to find $\sigma$, such that $\Gamma(\sigma,p)=\tau$.

We conclude that $P(\tr)$ is a~pseudomanifold, whose boundary is
a~union of the $(n-1)$-simplices $\tau=((W_0,G_0),\dots,(W_t,G_t))$,
such that $W_0\neq\es$, so then the subcomplex $\partial P(\tr)$
consists of all simplices $\sigma=((W_0,G_0),\dots,(W_t,G_t))$, such
that $G_0\neq\es$.  \end{proof}

\section{Immediate snapshot complexes as protocol complexes} \label{sect:6}

\subsection{The protocol complexes of a standard full-information protocol} 

\nin This section will provide a bridge between the mathematical and the
theoretical distributed computing contexts. Specifically, we shall
explain why immediate snapshot complexes provide a~correct
combinatorial model for the protocol complexes in the immediate
snapshot read/write computational model.

As in Section~\ref{sect:1}, assume that we have $n+1$ processes
indexed $0,\dots,n$, together with a~round counter
$\tr=(r_0,\dots,r_n)$. We consider the standard protocol associated to
this data. In this protocol, each process $p$ starts with some input
value $\alpha_p$, and then executes $r_p$ rounds. In each round, the
process $p$ first writes its current state into the register, which is
assigned to that process (full-information protocol), and then the
process reads the entire memory in one atomic step (snapshot read).

In the topological approach to distributed computing, once the
computational model is fixed, one associates a~simplicial complex to
each protocol. That complex is called a~{\it protocol complex}. We
refer to \cite{HKR} and the citations therein for the further
specifics of that construction. In general, the protocol complex is
defined as follows. The maximal simplices are indexed by all possible
executions of the protocol. The vertices of the protocol complexes are
the {\it local views} of individual processes. Two maximal simplices,
corresponding to executions $\sigma$ and $\tau$, share the simplex
consisting of those local views, which are the same in $\sigma$ and
in~$\tau$.

As was said above, the executions in the immediate snapshot read/write
computational model are shaped in layers. In each layer, a~group of
processes atomically writes to their respective registers, and then
takes an atomic snapshot of the entire memory. In other words, that
executions can be indexed by tuples $(W_1,\dots,W_t)$ of sets of
processes, where $W_1$ is the first group of processes which gets
activated, followed by $W_2$, and so on.

Let $Q(\tr)$ denote the protocol complex associated to the standard
full-information protocol for the round counter~$\tr$. In this case,
we have an additional condition $\sum_{i=1}^n\chi(p,W_i)=r_p$, for all
$p\in[n]$. Obviously, we have a~one-to-one correspondence between all
executions of the protocol and the top-dimensional simplices of the
immediate snapshot complex $P(\tr)$. To summarize, both $P(\tr)$ and
$Q(\tr)$ are pure of dimension $|\supp\tr|-1$, and we have a~natural
bijection between the sets of their top-dimensional simplices. Before
proceeding with extending this bijection, we need to analyze the
structure of information the processes write into the memory during
an~execution of the standard full-information protocol.

\subsection{Witness posets}
  
\nin When a~process is activated for the first time, the only
information that it has is its input value, so it will simply write
its input value into the assigned register. Later on, it will see the
information which other processes have written, and write that newly
acquired information, as a~part of his state, once it is activated
next time.  To describe this knowledge structure formally, let
$z_{p,k}$ denote the information which process~$p$ has {\it after} its
$k$th step (we cannot know for sure in which layer this step
takes place). Clearly, it is the same information as the one which
process~$p$ will {\it write} into the memory during its $(k+1)$th
step. For ease of notations, we set
$z_{p,0}:=\alpha_p$. Accordingly, $z_{p,r_p}$ denotes the information
which the process~$p$ has once it has executed the entire protocol. In
general, we shall write
\begin{equation}\label{eq:zpi}
z_{p,i}>z_{q,j}
\end{equation} 
to express the fact that {\it the process~$p$ after its $i$th
  step knows what the process~$q$ knew after its $j$th
  step.} Since all what processes learn during the execution of
the protocol is what other processes knew at various stages of the
execution (here we are thinking about the input values as the
knowledge of processes after the $0$th step) the entire
knowledge structure generated by an execution $\sigma$ of the protocol
is a poset, which we denote $Z(\sigma)$. This poset has elements
$z_{p,i}$, where $p\in[n]$ and $i\in[r_p]$, with the order relation
given by~\eqref{eq:zpi}.

\begin{df}\label{df:wip}
Assume $\tr=(r_0,\dots,r_n)$ is a~round counter, and $Z$ is a~poset,
whose set of elements is $\{z_{p,i}\,|\,p\in[n],i\in[k_p]\}$, for some
nonnegative integers $k_p\leq r_p$, for $p\in[n]$. For all $p\in[n]$,
$i\in[k_p]$, set $U(p,i):=Z_{<z_{p,i}}$, and set furthermore
$A(Z):=\{p\,|\,k_p=r_p\}$. The poset $Z$ is called a~{\bf witness
  poset} with parameter $\tr$ if its order relation satisfies the
following conditions:
\begin{enumerate}
\item[(1)] $z_{p,i+1}>z_{p,i}$, for all $p\in[n]$, $i\in[k_p-1]$;
\item[(2)] assume $p,q\in[n]$, $1\leq i\leq k_p$, and $1\leq j\leq
  k_q$, then one of the following three conditions is satisfied:
\begin{itemize}
\item  $U(q,j)\supset U(p,i)$, $z_{q,j}>z_{p,i-1}$, and $z_{p,i}\not >z_{q,j-1}$;
\item  $U(p,i)\supset U(q,j)$, $z_{p,i}>z_{q,j-1}$, and $z_{q,j}\not >z_{p,i-1}$;
\item  $U(q,j)=U(p,i)$, $z_{p,i}>z_{q,j-1}$, and $z_{q,j}>z_{p,i-1}$.
\end{itemize}
\item[(3)] the set of maximal elements of $Z$ is given by 
$\{z_{p,r_p}\,|\,p\in A(Z)\}$.
\end{enumerate}
We call $Z$ a~{\bf complete witness poset} if $A(Z)=[n]$.
\end{df}

\begin{figure}[hbt]
\centering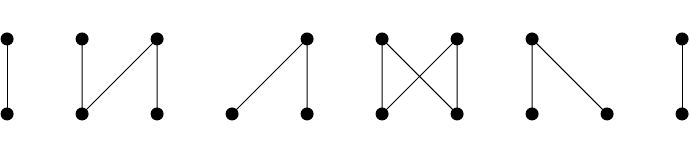
\caption{All witness posets for $\tr=(1,1)$.}
\label{fig:wp1}
\end{figure}

Note, that since a~witness poset $Z$ has to have some maximal
elements, there must exist $p$ such that $k_p=r_p$. Some examples of
witness posets are shown on Figure~\ref{fig:wp1}.

\begin{df}
Assume we are given a~round counter~$\tr$. We defined the simplicial 
complex $C(\tr)$ as follows:
\begin{itemize}
\item the set of vertices $V(C(\tr))$ consists of all witness posets
  $Z$ with parameter $\tr$, such that $|A(Z)|=1$;
\item a~subset $\{V_0,\dots,V_k\}\subseteq V(C(\tr))$ of vertices
  forms a~simplex if and only if there exists a~witness poset $Z$ with
  parameter $\tr$, such that 
\[\{I(Z,v)\,|\,v\in A(Z)\}=\ab\{V_0,\dots,V_k\},\]
where $I(Z,v)$ was defined in subsection~\ref{ssect:2.1}.
\end{itemize}
\end{df}

Note, that for a~witness poset $Z$ and $\es\neq B\subseteq A\subseteq
A(Z)$, we have
\[I(I(Z,A),B)=I(Z,B),\]
hence $C(\tr)$ is well-defined as a~simplicial complex. It is pure,
and has dimension $n$. The set of all simplices of $C(\tr)$ coincides
with the set of all witness posets with parameter $\tr$. The maximal
simplices of $C(\tr)$ are indexed by the complete witness posets.

\subsection{Protocol complexes vs witness posets} 

\nin To proceed, we need additional notation. Assume we have
a~sequence of sets $\sigma=(W_1,\ab\dots,\ab W_t)$, $p\in\cup_{i=1}^t
W_i$, and $k\in[t]$, we set
$M_\sigma(p,k):=\sum_{i=1}^k\chi(p,W_i)$. Furthermore, we let
$\rho_\sigma(p,k)$ denote the index such that $p$ occurs in
$W_{\rho_\sigma(p,k)}$ for the $k$th time. In other words,
$\rho_\sigma(p,k)$ is uniquely defined by the following two
conditions: $p\in W_{\rho_\sigma(p,k)}$ and
\begin{equation}\label{eq:mpk}
M_\sigma(p,\rho_\sigma(p,k))=k.
\end{equation} 
Note, that
\begin{equation}\label{eq:rho1}
\rho_\sigma(p,j)>\rho_\sigma(p,i)\textrm{ if } j>i,
\end{equation}
and 
\begin{equation}\label{eq:m1}
M_\sigma(p,k)\geq M_\sigma(p,l)\textrm{ if } k>l.
\end{equation}
Furthermore,
\begin{equation} \label{eq:rpm}
\textrm{ if } p\in W_k, \textrm{ then }\rho_\sigma(p,M_\sigma(p,k))=k.
\end{equation}
When $\sigma$ is clear from the context, we will skip it from the notations, 
and simply write $M(p,k)$ and $\rho(p,k)$.

\begin{prop}\label{prop:pr1}
For any round counter $\tr=(r_0,\dots,r_n)$, there is a~simplicial
isomorphism between the complexes $Q(\tr)$ and $C(\tr)$.
\end{prop}
\pr First, we define the map $\Phi$ which takes an execution
$\sigma=(W_1,\dots,W_t)$ of the protocol to a~complete witness poset
$Z=\Phi(\sigma)$. The set of the elements of $Z$ is taken to be
$\{z_{p,i}\,|\,p\in[n],i\in[r_p]\}$. The order relation is given by
the rule: for $p,q\in[n]$, $1\leq i\leq r_p$, $0\leq j\leq r_q$, we
have
\begin{equation} \label{eq:zphi}
z_{p,i}>z_{q,j}\textrm{ if and only if }\rho(p,i)\geq \rho(q,j+1).
\end{equation}
In words, the inequality~\eqref{eq:zphi} simply says that $q$ occurs
at least $j+1$ times in $W_1,\dots,W_k$, where $p$ occurs for the
$i$th time in $W_k$.

We check that $\Phi$ is well-defined. First, we check that $Z$ is
actually a~poset. Assume $z_{p,i}>z_{q,j}$ and
$z_{q,j}>z_{p,i}$. Then, \eqref{eq:zphi} implies that
$\rho(p,i)\geq\rho(q,j+1)$ and $\rho(q,j)\geq\rho(p,i+1)$. This gives
a~contradiction with~\eqref{eq:rho1}. Assume furthermore that
$z_{p,i}>z_{q,j}$ and $z_{q,j}>z_{s,k}$. Here, \eqref{eq:zphi} implies
that $\rho(p,i)\geq\rho(q,j+1)$ and
$\rho(q,j)\geq\rho(s,k+1)$. Using~\eqref{eq:rho1} we then conclude
that $\rho(p,i)>\rho(s,k+1)$, and hence $z_{p,i}>z_{s,k}$.

Second, we want to check that $Z$ is a complete witness poset, by
verifying the conditions in Definition~\ref{df:wip}. Condition~(1)
says that $z_{p,i+1}>z_{p,i}$, which \eqref{eq:zphi} translates to
$\rho(p,i+1)\geq\rho(p,i+1)$, which is a~tautology. 

Next, we check Condition~(2). We pick $p,q\in[n]$, $1\leq i\leq r_p$,
$1\leq j\leq r_p$, and compare $\rho(p,i)$ with $\rho(q,j)$.  Without
loss of generality, we can assume that $\rho(p,i)\geq\rho(q,j)$.  This
implies $z_{p,i}>z_{q,j-1}$. In addition, we can show that
$U(p,i)\supseteq U(q,j)$. Indeed, take $z_{s,k}<z_{q,j}$. By
\eqref{eq:zphi}, we have $\rho(q,j)\geq\rho(s,k+1)$. Since
$\rho(p,i)\geq\rho(q,j)$, we get $\rho(p,i)\geq\rho(s,k+1)$, and so
$z_{s,k}<z_{p,i}$. In particular, if $\rho(p,i)=\rho(q,j)$ then
repeating this argument gives $U(p,i)=U(q,j)$, $z_{p,i}>z_{q,j-1}$,
and $z_{q,j}>z_{p,i-1}$. On the other hand, if we have a~strict
inequality $\rho(p,i)>\rho(q,j)$ then $z_{p,i}>z_{q,j-1}$, and
$z_{q,j}\not>z_{p,i-1}$, which in turn implies that we have a~strict
inclusion $U(p,i)\supset U(q,j)$. In any case, Condition~(2) is
satisfied.

Finally, to check Condition~(3), as well the completeness, we need to
see that one cannot have $z_{p,r_p}>z_{q,r_q}$. This is so, since
otherwise we would have $M(q,\rho(p,r_p))\geq r_q+1$, which is
impossible.  We can therefore conclude that $\Phi(\sigma)$ is
a~well-defined complete witness poset.

Now, we define a map~$\Psi$, which takes an arbitrary complete witness
poset $Z$ with parameter $\tr$ to the protocol execution $\Psi(Z)$.
The condition that $U(p,i)$ is comparable with $U(q,j)$ for all
$p,q\in[n]$, $i\in[r_p]$, and $j\in[r_q]$, means that we can order all
$U(p,i)$'s by inclusion. So assume that for all $k=1,\dots,t$, we have
sets $S_k=\{(p_1^k,i_1^k),\ab\dots,\ab(p_{v_k}^k,i_{v_k}^k)\}$, such
that the following two conditions are true:
\begin{itemize}
\item $U(p_1^k,i_1^k)=\dots=U(p_{v_k}^k,i_{v_k}^k)$,  
\item $U(p_1^k,i_1^k)\subseteq U(p_1^{k+1},i_1^{k+1})$, for all
  $k=1,\dots,t-1$.
\end{itemize}
We set $W_k:=\{p_1^k,\dots,p_{v_k}^k\}$, for all
$k=1,\dots,t$. Clearly, $(W_1,\dots,W_t)$ is a well-defined execution,
since all indices in each $W_k$ are different, and the number of
occurrences of each $p$ is $r_p$.

Let now $\sigma=(W_1,\ab\dots,\ab W_t)$ be an execution of the
protocol and let us show that $\Psi\circ\Phi(\sigma)=\sigma$. First,
pick $p,q\in[n]$, and assume that $p,q\in W_k$ for some $1\leq k\leq
t$. By \eqref{eq:rpm}, we have
\[\rho(p,M(p,k))=\rho(q,M(q,k))=k.\] 
Assume $z_{s,i}<z_{p,M(p,k)}$, then by \eqref{eq:zphi}, this is
equivalent to $\rho(p,M(p,k))\geq\rho(s,i+1)$, which in turn is
equivalent to $\rho(q,M(q,k))\geq\rho(s,i+1)$, and hence to
$z_{s,i}<z_{q,M(q,k)}$.  This means that
\[U(p,M(p,k))=U(q,M(q,k)).\] 
Now, let us pick $p,q\in [n]$, and $l<k$, such that $p\in W_k$, $q\in
W_l$. Take $z_{s,i}<z_{q,M(q,l)}$, then
$l=\rho(q,M(q,l))\geq\rho(s,i+1)$.  Since $\rho(p,M(p,k))=k>l$, it
follows that $\rho(p,M(p,k))\geq\rho(s,i+1)$, hence
$z_{s,i}<z_{p,M(p,k)}$. Thus we see that in this case
\[U(p,M(p,k))\supset U(q,M(q,k)),\] 
where the inclusion is strict, since $z_{p,M(p,k)-1}\in
U(p,M(p,k))\setminus U(q,M(q,k))$. Together these two calculations
show that $\Psi\circ\Phi(\sigma)=\sigma$.

On the other hand, take an arbitrary complete witness poset $Z$ with
parameter $\tr$. Set $\wti Z:=\Phi\circ\Psi(Z)$. Note,
that both $Z$ and $\wti Z$ have the same sets of elements $z_{p,i}$,
for $p\in[n]$, $i\in[r_p]$, and we have $z_{p,i+1}>z_{p,i}$ for all
$p,i$ in both $Z$ and $\wti Z$. Assume now $z_{p,i}>z_{q,j}$ in $Z$.
By Definition~\ref{df:wip} this is equivalent to $U(p,i)\supseteq
U(q,j+1)$, which in turn is equivalent to $z_{p,i}>z_{q,j}$
in~$\widetilde Z$.

This shows that both $\Phi$ and $\Psi$ are well-defined and are
inverses of each other. Furthermore, since the information which the
process $p$ has after its $i$th run is precisely $U(p,i)$, the poset
$Z_{\leq z_{p,r_p}}$ is the local view of the process $p$, and taking
lower ideals corresponds to taking a~set of local views, which are
compatible in some execution. This means that $\Phi$ and $\Psi$ are
actually simplicial isomorphisms. 
\end{proof}

\subsection{Witness posets vs witness structures} 

\nin As a next step we show that witness posets and witness structures
encode identical simplicial information.

\begin{prop}\label{prop:pr2}
For any round counter $\tr=(r_0,\dots,r_n)$, we have a simplicial
isomorphism between complexes $C(\tr)$ and $P(\tr)$.
\end{prop}
\pr We describe maps $\wti\Phi:P(\tr)\rightarrow C(\tr)$ and
$\wti\Psi:C(\tr)\rightarrow P(\tr)$, which will generalize maps $\Phi$
and $\Psi$ from the proof of Proposition~\ref{prop:pr1}. Consider
a~witness structure $\sigma=\ab((W_0,G_0),\ab\dots,(W_t,G_t))$
indexing a~simplex of $P(\tr)$. Set $k_p:= \sum_{i=0}^t
\chi(p,W_i)-1$. Let the set of elements of $\wti\Phi(\sigma)$ be
$\{z_{p,i}\,|\,p\in [n],i\in[k_p]\}$. For $p\in W_0$, $k\in[k_p]$, 
we let $\tilde\rho(p,k)$ denote the index $\rho$, such that
$p\in W_\rho\cup G_\rho$ and $\sum_{i=1}^\rho\chi(p,W_i\cup
G_i)=k$. The order relation in $\wti\Phi(\sigma)$ is then given by: 
\[z_{p,i}>z_{q,j}\textrm{ if and only if }\tilde\rho(p,i)\geq\tilde\rho(q,j+1).\]
The verification that $\wti\Phi$ is well-defined is verbatim to that 
in Proposition~\ref{prop:pr2}. One also sees easily that 
$\wti\Phi(\sigma)=I(A(\sigma),\Phi(W_0\cup G_0,\dots,W_t\cup G_t))$.

Next, we define $\wti\Psi:C(\tr)\rightarrow P(\tr)$. Take $Z\in
C(\tr)$, and denote
$\wti\Psi(Z)=((W_0,G_0),\ab\dots,\ab(W_t,G_t))$. We let $W_0$ be given
by the identity $\min Z=\{z_{p,0}\,|\,p\in W_0\}$, and set
$G_0:=[n]\sm W_0$. Assume the set $S_1,\dots,S_t$ are chosen in the
same way as when we defined $\Psi$ in the proof of
Proposition~\ref{prop:pr2}, and set $U_k:=U(p_1^k,i_1^k)$, for all
$1\leq k\leq t$.  We have $U_1\subset\dots\subset U_t$, and we set
$W_k:=\{p_1^k,\dots,p_{v_k}^k\}$, for all $1\leq k\leq t$.  Assume we
have $p$ such that $k_p<r_p$. Let $m$ be the smallest index such that
$z_{p,k_p}\in U_m$, then $p\in G_m$; this index is well-defined, since
$z_{p,k_p}$ is not a~maximal element.  This rule defines uniquely the
sequence of sets $G_1,\dots,G_t$. The verification that $\wti\Psi$ is
well-defined is a~straightforward extension of the argument in the
proof of Proposition~\ref{prop:pr1}.

Same way as above, we can see that $\wti\Phi$ and $\wti\Psi$ are
inverses of each other. Furthermore, the map $\wti\Phi$ takes the
operation of taking lower ideals under a~subset of maximal elements to
the ghosting operation on the witness structures. It follows that
$\wti\Phi$ and $\wti\Psi$ are simplicial isomorphisms.  \end{proof}

\vskip5pt

\begin{figure}[hbt]
\centering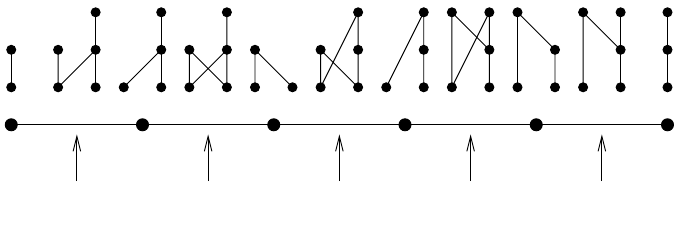
\caption{The immediate snapshot complex for $\tr=(2,1)$, with simplex
  names given as witness posets, as well as witness structures.}
\label{fig:wpws}
\end{figure}

Even though the information contained in simplicial complexes $C(\tr)$
and $P(\tr)$ is the same, in various situations it can be more
convenient to use one or the other. We feel that taking lower ideals
is simpler to grasp than the ghosting operation. On the other hand,
the entire witness poset structure is a bit awkward to describe, when
we want to work with specific simplices, here, witness structures
provide a~more succinct description. Figure~\ref{fig:wpws} shows the
parallel combinatorial encodings of the simplices in the simple case
$\tr=(2,1)$.

\section*{Acknowledgments}
\nin The author would like to thank Maurice Herlihy for useful
discussions of the immediate snapshot computational models, and Demet
Taylan for finding several typos. The research was supported by DFG
grant as a part of University of Bremen Excellence Initiative.


\end{document}